\documentclass[DIV=calc, paper=letter, fontsize=11pt]{scrartcl}	 

\usepackage[]{units}
\usepackage{lipsum} 
\usepackage[english]{babel} 
\usepackage[protrusion=true,expansion=true]{microtype} 
\usepackage{amsmath,amssymb,amsfonts,amsthm} 
\usepackage[svgnames, table]{xcolor} 
\usepackage[hang, small,labelfont=bf,up,textfont=it,up]{caption} 
\usepackage{booktabs} 
\usepackage{fix-cm}	 
\usepackage[]{units}

\usepackage{tabularx} 

\usepackage[colorlinks=true, linkcolor=blue, citecolor=blue, filecolor=blue, runcolor=blue, urlcolor = blue]{hyperref}

\usepackage{fullpage}
\usepackage{sectsty} 

\usepackage{fancyhdr} 
\usepackage{lastpage} 

\usepackage{nicefrac}
\usepackage{cite}

\usepackage{tikz}
\usetikzlibrary{calc,patterns,decorations.pathmorphing,decorations.markings}
\usepackage{pgfplots}

\usepackage{multicol}
\usepackage{float} 

\usepackage{subcaption}

\newcounter{tempcolnum}
\makeatletter
\newcommand{\multicolinterrupt}[1]{
\setcounter{tempcolnum}{\col@number}
\end{multicols}
#1%
\begin{multicols}{\value{tempcolnum}}
}
\makeatother

\usepackage{graphicx}
\DeclareGraphicsExtensions{.png}

\usepackage{lettrine} 
\newcommand{\initial}[1]{ 
\lettrine[lines=3,lhang=0.3,nindent=0em]{
\color{DarkGoldenrod}
{\textsf{#1}}}{}}

\usepackage{caption}

\usepackage{graphicx}

\usepackage{titling} 

\newcommand{\HorRule}{\color{DarkGoldenrod} \rule{\linewidth}{1pt}} 

\pretitle{\vspace{-80pt} \begin{flushleft} \HorRule \fontsize{20}{20} \color{DarkRed} \selectfont} 

\title{Hydrodynamic Flows on Curved Surfaces: Spectral Numerical Methods for Radial Manifold Shapes} 

\posttitle{\par\end{flushleft}} 
\preauthor{\vspace{-5pt} \begin{flushleft}\fontsize{14}{14} \large  \color{DarkRed}} 
\author{B. J. Gross and P. J. Atzberger$^{*}$ } 
\postauthor{\fontsize{12}{12} \small \color{Black} 
Department of Mathematics, Department of Mechanical Engineering, University of California Santa Barbara; atzberg@gmail.com; http://atzberger.org/ \\
\vskip 1.5em
\lfoot{\footnotesize * Work supported by DOE ASCR CM4 DE-SC0009254, NSF CAREER Grant DMS-0956210, \\ and NSF Grant DMS - 1616353.}
\end{flushleft} \vspace{-18pt} \HorRule} 
\date{\today}  

\newcommand{\mb}[1]{\mathbf{#1}}
\newcommand{\bs}[1]{\boldsymbol{#1}}

\newcommand{\mDiv}{ {\mbox{div}} }
\newcommand{\mGrad}{ {\mbox{grad}} }
\newcommand{\mCurl}{ {\mbox{curl}} }

\DeclareGraphicsExtensions{.png}

\begin{document}

\maketitle 

\thispagestyle{fancy} 


\initial{W}\textbf{e formulate hydrodynamic equations and spectrally accurate numerical methods for investigating the role of geometry in flows within two-dimensional fluid interfaces. To achieve numerical approximations having high precision and level of symmetry for radial manifold shapes, we develop spectral Galerkin methods based on hyperinterpolation with Lebedev quadratures for $L^2$-projection to spherical harmonics.  We demonstrate our methods by investigating hydrodynamic responses as the surface geometry is varied.  Relative to the case of a sphere, we find significant changes can occur in the observed hydrodynamic flow responses as exhibited by quantitative and topological transitions in the structure of the flow.  We present numerical results based on the Rayleigh-Dissipation principle to gain further insights into these flow responses.  We investigate the roles played by the geometry especially concerning the positive and negative Gaussian curvature of the interface.  We provide general approaches for taking geometric effects into account for investigations of hydrodynamic phenomena within curved fluid interfaces.
}

\setlength{\parindent}{5ex}

\section{Introduction}
\label{sec_intro}
We develop spectral numerical methods for a continuum mechanics formulation of hydrodynamic flows within two-dimensional curved fluid interfaces.  Hydrodynamics within curved geometries play an important role in diverse physical systems including the thin films of soap bubbles~\cite{Kellay2017,Kornek2010,LealFeng1997,bookKimMicroHydro1991}, lipid bilayer membranes~\cite{HonerkampSmith2013,AtzbergerBassereau2014,AtzbergerSoftMatter2016,
Saffman1975,Powers2002} and recent interface-embedded colloidal systems~\cite{StebePNAS2011,Choi2011,Ershov2013,Bresme2007,DonevInterface2018,Bresme2007}.  Similar hydrodynamic and related curvature mediated phenomena also plays an important role in physiology such as in the cornea of the eye with its tear film~\cite{Braun2012}, transport of surfactants in lung alveoli~\cite{Hermans2015,SquiresManikantan2017} or in cell mechanics~\cite{Mogilner2018,Chou2010,Powers2002}.  Each of these systems involve potential interactions between the curvature of the interface and hydrodynamic flows.  We investigate these types of flows by formulating continuum mechanics equations for hydrodynamics using variational principles and the exterior calculus of differential geometry~\cite{Marsden1994}.  This provides an abstraction that is helpful in generalizing many of the techniques of fluid mechanics to the manifold setting while avoiding many of the tedious coordinate-based calculations of tensor calculus.  The exterior calculus formulation also provides a coordinate-invariant set of equations helpful in providing insights into the roles played by the geometry in the hydrodynamics.  

{There has been a significant amount of experimental and theoretical work developing approaches for investigating hydrodynamics within curved fluid interfaces~\cite{Pozrikidis2001a,Powers2010a,Kralchevsky1994,Scriven1960}. Experimental work includes single particle tracking of inclusions to determine information about interfacial viscosity and diffusivities~\cite{AtzbergerBassereau2014,Woodhouse2012}.}  Even the formulation of the correct continuum mechanics equations presents some significant challenges in the manifold setting~\cite{Marsden1994}.  For instance in a two-dimensional curved fluid sheet the equations must account for the distinct components of linear momentum correctly.  The concept of momentum is not an intrinsic field of the manifold and must be interpreted with respect to the ambient physical space~\cite{Marsden1994}.  For instance, when considering non-relativistic mechanics in an inertial reference frame with coordinates $x,y,z$, the $x$-component of momentum is a conserved quantity distinct from the $y$ and $z$-components of momentum.  An early derivation using coordinate-based tensor calculus in the ambient space was given for hydrodynamics within a curved two-dimensional fluid interface in Scriven~\cite{Scriven1960}.  This was based on the more general shell theories developed in~\cite{Chien1942,Ericksen1957}.  Many subsequent derivations have been performed using tensor calculus for related fluid-elastic interfaces motivated by applications.  This includes derivation of equations for surface rheology~\cite{EdwardsRheologyIPart2Y1988,HowardPart1Y1993,
HowardPart2Y1993}, investigation of red-blood cells~\cite{Secomb1982}, surface transport in capsules and surfactants on bubbles~\cite{Pozrikidis2001a,LealFeng1997}, and  investigations of the mechanics, diffusion, and fluctuations associated with curved lipid bilayer membranes~\cite{Rahimi2013a,Steigmann1999a,LubenskyMemDynamicsFluct1995,
Henle2008,Vlahovska2007,SteigmiannOster2013, 
DesernoDiffGeo2015,
AtzbergerSoftMatter2016,
SahuSauerMandadapu2017}.
Recent works by Marsden et al.~\cite{Marsden1994,Marsden2007,MarsdenOrtiz2006} develop the continuum mechanics in the more general setting when both the reference body and ambient space are treated as general manifolds as the basis for rigorous foundations for elasticity~\cite{Marsden1994,Marsden2007}.  In this work, some of the challenges associated with momentum and stress with reference to ambient space can be further abstracted in calculations by the use of covector-valued differential forms and a generalized mixed type of divergence operator~\cite{Marsden2007,MarsdenOrtiz2006}.  A particularly appealing way to derive the conservation laws for manifolds is through the use of variational principles based on the balance of energy and symmetries~\cite{Marsden1994}.  This has recently been pursued to derive elastic and hydrodynamic equations for lipid membranes in~\cite{Guven2018,ArroyoRelaxationDynamics2009,
SahuSauerMandadapu2017}.  We briefly present related derivations based on the energy balance approach of~\cite{Marsden1994,MarsdenOrtiz2006} to obtain our hydrodynamic equations in Section~\ref{sec_cont_mechanics}.

\lfoot{} 

There has been a lot of recent interest and work on developing computational methods for evaluating differential operators and for solving equations on curved surfaces~\cite{BertozziFourthOrderGeometries2006,
SchroderDesbrunSimplicialFluids2007,KeenanSiggraph2013,Sethian2013}.  This has been motivated in part by applications in computer graphics~\cite{KeenanSiggraph2013,
DesbrunHirani2003,Azencot2015} and interest in applications using shell theories for elasticity and hydrodynamics~\cite{ArroyoRelaxationDynamics2009,
Guven2018,DesernoMullerParticleInterface2005,
DesernoDiffGeo2015,Powers2002,SauerDuongMandadapuFEM2017,
Yavari2008}.  Many computational methods treat the geometry using a triangulated surface and build discrete operators to model their curvature and differential counter-parts~\cite{LinTriangulation1982,
McIvor1997,Stokely1992,
Taubin1995,Desbrun2003,Zorin2005,AtzbergerPeskinOsmotic2015}.  Some early work in this direction includes~\cite{LinTriangulation1982,McIvor1997,Stokely1992,Taubin1995} and the Surface Evolver of Brakke~\cite{brakkeSurfaceEvolver1992}.  More recently, discrete approaches such as the Discrete Exterior Calculus (DEC)~\cite{Desbrun2003,SchroderDEC2005}, Finite Element Exterior Calculus (FEEC)~\cite{Arnold2006,HolstGilette2017} and Mimetic Methods (MM)~\cite{Bochev2006} have been developed that aim to reproduce in the numerics analogous properties of the differential operators related to the geometric and topological structure of the manifold~\cite{Bochev2006,Desbrun2003,SchroderDEC2005,Zorin2005,
Arnold2006,HolstGilette2017,Hirani2003}.  For manifolds represented by discrete symplicial complexes this includes preserving the adjoint conditions between the exterior derivative, boundary operator, and co-differentials to create a discrete analogue of the de Rham complex and related theory~\cite{Abraham1988,SpivakDiffGeo1999,JostBookDiffGeo1991}.  This has been used to obtain models of surface Laplacians and results like discrete Hodge decompositions~\cite{Bochev2006,Arnold2006,HolstGilette2017,KeenanSiggraph2013}.  For finite elements these properties can be shown to be essential for discretizing problems in elasticity and fluid mechanics to obtain well-posed approximations with stable numerical methods~\cite{Bochev2006,Arnold2006,HolstGilette2017}.  In the DEC approach to formulating numerical methods for PDEs on manifolds, the methods obtained are similar to finite differences~\cite{Desbrun2003,Hirani2003}.  This work has allowed for impressive results including schemes that are exactly conservative for quantities such as mass and vorticity~\cite{HiraniNavierStokes2016,SchroderDesbrunSimplicialFluids2007}.  Deriving operators preserving geometric structure is non-trivial and current numerical methods for spherical topologies are typically first or second order accurate~\cite{HiraniNavierStokes2016,SchroderDesbrunSimplicialFluids2007}.  Recent methods have been developed in the setting of tensor product basis that are spectrally accurate in~\cite{DesbrunSpectralChainCollocation2014}.

Here we develop spectrally accurate methods for solving hydrodynamic flows on curved surfaces having general radial manifold shape based on an exterior calculus formulation of the hydrodynamic equations.  {While our numerical methods do not seek to preserve exactly the geometric and topological relations between our approximate exterior operators, we have from the spectral representation that these relations hold for our expansions to a high level of accuracy.}  In our derivations we make use of the relations in the exterior calculus such as the Hodge decomposition and adjoint conditions on exterior derivatives and co-differentials to obtain our weak approximations to the hydrodynamic equations.  

We develop spectral numerical methods based on Galerkin approximations with hyperinterpolation for $L^2$-projection to spherical harmonics based on Lebedev quadrature.  The Lebedev quadrature nodes are derived by solving a non-linear system of equations that impose both exactness of integration on spherical harmonics up to a specified order while maintaining symmetry under octahedral rotations and reflections~\cite{Lebedev1976,Lebedev1999}.  While one could also consider using a quadrature based on spherical coordinates and sampling on the latitudinal and longitudinal points which have some computational advantages by using the Fast Fourier Transform~\cite{HealyDriscoll1994,Kunis2003,Healy2003}, as we discuss in Section~\ref{sec_surf_quad}, these nodes have significant asymmetries with nodes concentrated in clusters near the poles of the sphere.  Since Lebedev nodes were developed for quadratures on the sphere, we extend them to obtain quadratures for general radial manifolds by making use of a coordinate-independent change of measure formula derived using the Radon-Nikodym Theorem~\cite{Lieb2001}.  {We mention that alternative formulations are also possible related to our approach in terms of discrete triangulations provided appropriate transport theorems and quadrature are developed over the mesh.}  We test the accuracy of our quadrature scheme for general radial manifolds by integrating the Gaussian curvature over the surface.  From the Gauss-Bonnet Theorem this should give the Euler characteristic for the spherical topology and independent of the detailed geometric shape~\cite{Pressley2001,SpivakDiffGeo1999}.  We then provide convergence results for our hydrodynamics solver in a few special cases with known hydrodynamics solutions showing the solver's accuracy.

We demonstrate our numerical methods for a few example manifolds by investigating hydrodynamic flow responses and the role of surface geometry.  As a baseline, we first consider hydrodynamic flows driven by particles configured on a sphere and subjected to force.  We investigate for equivalent forcing how these hydrodynamic flow responses change when the particles are on a surface having a more complicated geometry.  We find as the geometry has more heterogeneous curvatures the structure of the hydrodynamic flow fields are observed to change significantly.  In some cases this appears to correspond to a topological transformation of the stream-lines of the flow.  The fluid flow is observed to result in a transition from a more global recirculation of fluid to a much more localized recirculation of fluid and can result in the creation of new singularities (stagnation points) that take the form of new vortices and saddle points.  We use our numerical methods to report on the Rayleigh-Dissipation associated with these hydrodynamic flow responses.  We find that these quantitative and topological changes may play a role in mitigating dissipation within the fluid for certain geometries.  We mention that related to our findings some work on fluid dissipation in the case of perturbations from a flat sheet, such as in a planar soap film, were recently investigated in~\cite{Debus2017}.  In summary, our methods and results show in the context of radial manifolds some of the rich ways that surface geometry can impact hydrodynamic flow responses within curved fluid interfaces.

We have organized our paper as follows.  We discuss briefly the formulation of the conservation laws of continuum mechanics in the manifold setting in Section~\ref{sec_cont_mechanics}.  We formulate our hydrodynamic equations for curved fluid interfaces in Section~\ref{sec:hydro_formulate}.   We develop a Hodge decomposition on manifolds for the fluid velocity field in Section~\ref{sec_hodge_decomp}.  We develop our spectral numerical methods and discuss hyperinterpolation for $L^2$-projection to spherical harmonics based on Lebedev quadrature in Section~\ref{sec_num_methods}.  We extend the Lebedev quadrature to general radial manifolds in Section~\ref{sec_surf_quad}.  We provide convergence results for our quadratures in Section~\ref{sec_quad_validation}.  We discuss our formulation of the Galerkin approximation based on our Hodge decomposition of the fluid velocity in Section~\ref{sec_Galerkin_approx}.  We then present convergence results for our hydrodynamics solver in Section~\ref{sec_convergence}.  Finally, we demonstrate our methods by investigating hydrodynamic flow responses for a few radial manifold shapes in Section~\ref{sec_results_hydro_flows}.  We investigate how hydrodynamic responses change as the geometry is varied.  We present results showing that significant changes can occur in the quantitative and topological structure of the observed hydrodynamic flow responses.  We also present results using Rayleigh-Dissipation rates to characterize these flows.  We then summarize our methods and discuss some of the ways that our approaches could be useful for solving other PDEs on manifolds and performing further investigations.  We expect our numerical methods to be useful quite generally for developing further techniques and simulation methods for investigating  hydrodynamic phenomena within curved fluid interfaces.

\section{Continuum Mechanics for Manifolds}
\label{sec_cont_mechanics}

\begin{figure}[H]
\centering
\includegraphics[width=0.99\linewidth]{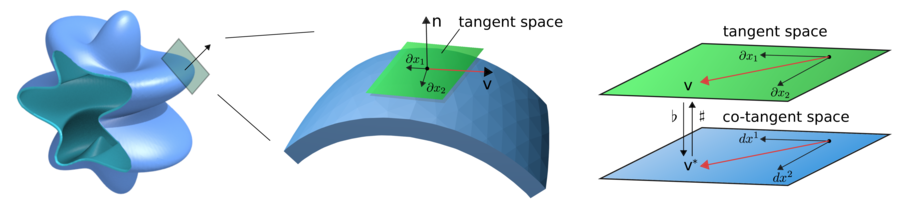}
\caption{Continuum Mechanics for Manifolds.  We consider continuum mechanics formulated within general curved spaces having non-euclidean metric.  At each point of the manifold the tangent space $T\mathcal{M}_{\mb{x}}$ consists of all vectors tangent to the manifold when thought of as an embedding.  The cotangent space $T^{*}\mathcal{M}_{\mb{x}}$ consists of all linear functionals of the tangent vectors giving a dual space.  Elements of these spaces are isomorphic and can be related by the operators $\flat: T\mathcal{M}_{\mb{x}} \rightarrow T^*\mathcal{M}_{\mb{x}}$ and $\sharp: T\mathcal{M}_{\mb{x}} \rightarrow T^*\mathcal{M}_{\mb{x}}$ see discussion in Section~\ref{sec_exterior_calc}.}
\label{fig:manifold_cont_mech}
\end{figure}

We discuss briefly how we formulate continuum mechanics in the covariant form on curved surfaces and more general manifolds as in Figure~\ref{fig:manifold_cont_mech}.  Our approach makes use of exterior calculus approaches~\cite{Abraham1988,SpivakDiffGeo1999}.  By using an exterior calculus formulation of the hydrodynamics we can abstract away in our derivations and analysis many of the details related to intricacies of computing with specific coordinates and tensors in curved spaces.  This allows us to generalize more readily many of the techniques employed in fluid mechanics to the manifold setting.  This approach also helps in revealing geometric features of the equations and the continuum mechanics.  We first discuss very briefly some background on exterior calculus and then discuss the derivations of our hydrodynamic equations for curved fluid interfaces.

\subsection{Differential Geometry and Exterior Calculus}
\label{sec_exterior_calc}
The exterior calculus allows for a covariant formulation of the equations avoiding the need to explicitly express the metric tensor and components.  This reveals more readily and intuitively in many cases the roles played by geometry.  The exterior calculus will ultimately simplify many of our calculations in fluid mechanics by providing an analogue to often employed vector calculus techniques used in the Euclidean case.  We briefly review for exterior calculus the basic definitions and relations we shall make use of in our calculations.  For a more comprehensive and detailed discussion see~\cite{Abraham1988,SpivakDiffGeo1999}.  

We consider smooth $n$-dimensional closed Riemann manifolds $\mathcal{M}$ with metric $\mb{g}$.  We have from the Whitney Embedding Theorem~\cite{Abraham1988,Whitney1936} that we can always express such manifolds as an embedding in a space $\mathbb{R}^N$ provided $N$ is sufficiently large.  We consider the embedding map $\bs{\sigma} = \bs{\sigma}(\mb{x}) : \mathbb{R}^n \rightarrow \mathbb{R}^N$ associated with a chart having coordinates $x^i$.  The tangent space at location $\mb{x}$ consists of the span of the vectors $\partial \bs{\sigma}/\partial_{x^i} \in \mathbb{R}^N$.  We denote the tangent space by $T\mathcal{M}_\mb{x} = \mbox{span}\left\{ \partial_{x^i}\right\}$ with the usual notation for the basis vectors $\partial_{x^i} := \partial\bs{\sigma}/\partial_{x^i}=\bs{\sigma}_{x^i}$~\cite{Abraham1988}.  The co-tangent space corresponds to the dual $T^*\mathcal{M}_\mb{x}$ of the tangent space consisting of all linear functionals acting on vectors in $T\mathcal{M}_\mb{x}$.  The linear functional corresponding to vector $\mb{u} = u^{i} \partial_{x^i}$ is denoted by $\mb{u}^*$ having the action on a vector $\mb{v} = v^i \partial_{x^i}$ given by $\mb{u}^*[\mb{v}] = u^ig_{ij}v^j$ where $g_{ij} = \bs{\sigma}_{x^i}\cdot\bs{\sigma}_{x^j} = \langle\partial_{x^i},\partial_{x^j} \rangle$ is the metric tensor.  We can use as a basis for the co-tangent space $\mb{dx}^i$ which denotes the linear functional having the action $\mb{dx}^i[\partial_{x^j}] = \delta_{j}^i$ where $\delta_{j}^i$ is the Kronecker $\delta$-function which is one if $i=j$ and otherwise zero.  This allows for representing a general linear functional of the co-tangent space as $\mb{u}^* = u_i \mb{d}x^i$.  With these conventions we can define isomorphisms between the tangent and co-tangent spaces $\flat: T\mathcal{M}_\mb{x} \rightarrow T^*\mathcal{M}_\mb{x}$ and $\sharp: T^*\mathcal{M}_\mb{x} \rightarrow T\mathcal{M}_\mb{x}$.  These correspond to the relationships between the representations $\mb{u} = u^{i} \partial_{x^i}$ and $\mb{u}^{*}= u_i \mb{d}x^i$ with $\mb{u}^{\flat} = \mb{u}^*$ and $\lbrack \mb{u}^*\rbrack^{\sharp} = \mb{u}$.  In terms of coordinates these maps give $[\mb{u}^{\flat}]_i = u_i = g_{ij}u^i$ and $[\mb{u}^{\sharp}]^i = u^i = g^{ij}u_j$.  The $g_{ij}$ are the components of the metric tensor and $g^{ij}$ are components of the inverse metric tensor.  The $\sharp$ and $\flat$ are often referred to as the musical isomorphisms since this helps remember how they correspond in coordinates to lowering and raising the indices~\cite{Abraham1988}.

We can also define functionals that take multiple vectors as input.  We define $k$-forms $\bs{\alpha}(\mb{u}_1, \ldots, \mb{u}_k)$ as functionals that are linear and have the anti-symmetric property $\bs{\alpha}(\mb{u}_{\sigma(1)}, \ldots, \mb{u}_{\sigma(k)}) = \mbox{sign}(\sigma)\bs{\alpha}(\mb{u}_1, \ldots, \mb{u}_k)$.  We can combine $k$-forms $\bs{\alpha}$ and $\ell$-forms $\bs{\beta}$ to obtain $(k + \ell)$-forms using the wedge product $\wedge$ defined as 
\begin{eqnarray}
\bs{\alpha} \wedge \bs{\beta} = 
\frac{1}{k!\ell !} 
\sum_{\sigma \in \mathbb{P}_{k + \ell}} \mbox{sign}(\sigma) \bs{\alpha}(\mb{u}_{\sigma(1)},\ldots,\bs{u}_{\sigma(k)})\bs{\beta}(\mb{u}_{\sigma(k+1)},\ldots,\bs{u}_{\sigma(k+\ell)}).
\end{eqnarray}
The $\mathbb{P}_{k + \ell}$ is the permutation group on $k+\ell$ elements. This has the useful property that 
$\bs{\alpha} \wedge \bs{\beta} = (-1)^{k\ell} \bs{\beta} \wedge \bs{\alpha}$.  This allows us to construct from the $1$-forms $\bs{\alpha}$, $\bs{\beta}$ the differential forms that can be integrated over a two-dimensional surface as $\bs{\lambda} = \bs{\alpha}\wedge \bs{\beta}$ or more generally $n$-dimensional surfaces as $\bs{\lambda} = \bs{\alpha}^{(1)}\wedge \cdots \wedge \bs{\alpha}^{(n)}$.  We define the exterior derivative $\mb{d}$ acting on a $k$-form 
$\bs{\alpha} = \alpha_{i_1,\ldots, i_k} \mb{d}\mb{x}^{i_1}\wedge \cdots \mb{d}\mb{x}^{i_k}$ as the resulting $(k+1)$-form
\begin{eqnarray}
\mb{d}\bs{\alpha} = \frac{1}{k!}\frac{\partial}{\partial x^j} \alpha_{i_1,\ldots, i_k} \mb{d}\mb{x}^{j} \wedge \mb{d}\mb{x}^{i_1}\wedge \cdots \mb{d}\mb{x}^{i_k}.
\end{eqnarray}
When integrating over the manifold it is useful to define an $L^2$-inner-product on differential forms.  We use the volume $n$-form of the manifold $\omega$ to define the manifold $L^2$-inner-product
\begin{eqnarray}
\label{equ_manifold_inner_prod}
\left \langle \bs{\alpha},\bs{\beta} \right \rangle_{\mathcal{M}}
= 
\int_{\mathcal{M}}  \left \langle \bs{\alpha},\bs{\beta} \right \rangle_{g} \omega.
\end{eqnarray}
By using the isomorphisms $\sharp$ and $\flat$ between the co-tangent and tangent space we have that the local metric inner-product on differential forms is equal to the local metric inner-product on the isomorphic vector fields $\left \langle \bs{\alpha},\bs{\beta} \right \rangle_{g} = \left \langle \bs{\alpha}^{\sharp},\bs{\beta}^{\sharp} \right \rangle_{g}$.  As a consequence, the $L^2$-inner-product on differential forms is consistent with the $L^2$-inner-product defined on the isomorphic vector fields 
\begin{eqnarray}
\left \langle \bs{\alpha},\bs{\beta} \right \rangle_{\mathcal{M}} = 
\left \langle \bs{\alpha}^{\sharp},\bs{\beta}^{\sharp} \right \rangle_{\mathcal{M}}.
\end{eqnarray}
This is useful allowing us to perform calculations in either representation as convenient.  We define the Hodge star $\star$ as the operator that for any $k$-form $\bs{\gamma}$ and $\bs{\lambda}$ gives 
\begin{eqnarray}
\int_{\mathcal{M}} \bs{\gamma}\wedge \star \bs{\lambda} = \int_{\mathcal{M}} \langle \bs{\gamma}, \bs{\lambda} \rangle_{g} \omega = \langle \bs{\gamma}, \bs{\lambda} \rangle_{\mathcal{M}}.
\end{eqnarray}
where $\omega$ is the volume $n$-form of the closed manifold.  In coordinates we can express the Hodge star $\star\bs{\alpha}$ as the $(n-k)$-form 
\begin{eqnarray}
\star \bs{\alpha} = \frac{\sqrt{|g|}}{(n-k)!k!} \alpha^{i_1,\ldots,i_k} \epsilon_{i_1,\ldots,i_k,j_1,\ldots,j_{n-k}} \mb{d}x^{j_1}\wedge \cdots \wedge \mb{d}x^{j_{n-k}}.
\end{eqnarray} 
The $\epsilon_{\ell_1,\ldots,\ell_n}$ denotes the Levi-Civita tensor which gives the sign of the permutation of the indices $\ell_1,\ldots,\ell_n$ and is otherwise zero.  When working with differential forms the Hodge star gives the analogue for differential forms of taking the orthogonal complement of a vector subspace~\cite{Abraham1988}.  

For smooth orientable $n$-dimensional manifolds the Hodge star satisfies a number of useful identities that we shall use in our derivations.  The first is that $\star \star \bs{\eta} = (-1)^{k(n-k)} \bs{\eta}$ where $\bs{\eta}$ is a $k$-form.  If we have $\bs{\gamma} = \star \bs{\eta}$ then we can use two applications of the Hodge-star to determine that the inverse is again a Hodge star operator multiplied by a sign so that $\bs{\eta} = \star^{-1} \bs{\gamma} = (-1)^{k(n-k)} \star \bs{\gamma}$.  From these relations we have the useful identity that
\begin{eqnarray}
\langle \bs{\gamma}, \bs{\lambda} \rangle_{\mathcal{M}}
= 
\langle \star \bs{\gamma}, \star \bs{\lambda} \rangle_{\mathcal{M}}.
\end{eqnarray}
Furthermore, we have the adjoint relationship 
\begin{eqnarray}
\langle \star \bs{\gamma}, \bs{\lambda} \rangle_{\mathcal{M}}
= 
(-1)^{k(n-k)} \langle \bs{\gamma}, \star \bs{\lambda} \rangle_{\mathcal{M}}.
\end{eqnarray}
In other words, the adjoint operator to the Hodge star is $\star^T = (-1)^{k(n-k)}\star$.  It is often useful to work with the adjoint of the exterior derivative $\bs{\delta} = \mb{d}^T$ where the co-differential operator is defined as $\bs{\delta} = (-1)^{n(k-1) + 1} \star \mb{d} \star$.  With these definitions it follows that the exterior derivative $\mb{d}$ and codifferential $\bs{\delta}$ satisfy
\begin{eqnarray}
\label{equ_adjoint_d_delta}
\langle \bs{\delta} \bs{\alpha}, \bs{\beta} \rangle_{\mathcal{M}}
= 
\langle \bs{\alpha}, \mb{d} \bs{\beta} \rangle_{\mathcal{M}}.
\end{eqnarray}
These summarize a few relationships and identities in exterior calculus that we shall make use of in our subsequent calculations.  With these conventions we can generalize many of the operators that arise in vector calculus in the context of surfaces as  
\begin{eqnarray}
\label{equ:gradDivCur}
\mGrad_{\mathcal{M}}(f) = \lbrack \mb{d}f\rbrack^{\sharp}, \hspace{1cm}
\mDiv_{\mathcal{M}}(\mb{F}) =  -(-\star \mb{d}\star \mb{F}^\flat) = -\bs{\delta} \mb{F}^\flat, \hspace{1cm}
\mCurl_{\mathcal{M}}(\mb{F}) = \left\lbrack -\star \mb{d}\mb{F}^\flat \right\rbrack^{\sharp}.
\end{eqnarray}
The $f$ is a function ($0$-form) and $\mb{F}$ is a $1$-form.  For a more detailed discussion of exterior calculus see~\cite{Abraham1988,SpivakDiffGeo1999}.  

When working with mechanical systems involving manifolds embedded within an ambient space and when working with coordinates it is useful to summarize briefly a few definitions and results from differential geometry.  A more detailed discussion can be found in~\cite{Marsden1994,Abraham1988}.  When working in coordinates in an embedding space, the Christoffel symbols $\Gamma_{ab}^c$ of the manifold can be viewed as serving to represent the rate-of-change of the tangent space basis vectors $\mb{e}_a = \partial_{x^a}$ as $\frac{\partial \mb{e}_a}{\partial {x^b}} = \Gamma_{ab}^c \mb{e}_c$.  A more geometrically intrinsic definition without reference to an embedding space can be given in terms of the metric as 
\begin{eqnarray}
\label{equ_christoffel}
\Gamma_{ab}^c = \frac{1}{2} g^{ck}\left(\frac{\partial g_{ak}}{\partial x^b} + \frac{\partial g_{kb}}{\partial x^a} - \frac{\partial g_{ab}}{\partial x^k}\right).
\end{eqnarray}
We can define a covariant derivative connecting tangent spaces as 
\begin{eqnarray}
\label{equ_cov_deriv}
\nabla_{\mb{v}} \mb{w} = \left(\frac{\partial w^c}{\partial x^b}v^{b} + \Gamma_{ab}^c v^{a}w^{b}\right) \mb{e}_c.
\end{eqnarray}
When $\mb{w} = w^a\mb{e}_a = w^a\partial_{x^a}$ we denote this in components as $w^c_{|b}$ so that $\nabla_{\mb{e}_b} \mb{w} = w^c_{|b} \mb{e}_c$.  We can define a covariant divergence as $\overline{\mbox{div}}(\mb{w}) = w^b_{|b}$.  We often will consider motions $\phi_t : \mathcal{B} \rightarrow \mathcal{S}$ of a manifold with reference body $\mathcal{B}$ mapped to a configuration in an embedding space $\mathcal{S}$. We define the flow $\phi_{t,s}$ of a motion starting at time $t$ and ending at time $s$ as $\phi_{t,s} = \phi_{s}\circ \phi^{-1}_{t}$.  The associated velocity of the flow is given by a velocity field $\mb{v}$ with $\mb{v}
(\mb{x}) = \left.\frac{d}{ds} (\phi_{t,s} \mb{x}) \right|_{s = t}$. 
The Lie derivative of a tensor under the flow is defined as 
\begin{eqnarray}
\label{equ_Lie_deriv_abstract}
L_{\mb{v}} \mb{t} = \frac{d}{ds} \left. \left( \phi^{*}_{t,s}   \mb{t}\right) \right|_{s = t}.
\end{eqnarray}
The $\phi^{*}_{t,s}$ is the pull-back~\cite{Abraham1988}.   In the case of the tensor $\mb{t}$ is a vector field $\mb{t} = \mb{w}$ we have that the Lie derivative generalizes the material derivative as 
\begin{eqnarray}
\label{equ_Lie_deriv_vec}
\frac{D\mb{w}}{Dt} := 
L_{\mb{v}} \mb{w} = \partial_t \mb{w} + \nabla_{\mb{v}} \mb{w}.
\end{eqnarray}
We emphasize that $\nabla_{\mb{v}}$ is now the covariant derivative of equation~\ref{equ_cov_deriv}.
We also discuss additional results and relations in our prior work~\cite{AtzbergerGross2017} and Appendix~\ref{appendix:radialManifold} and~\ref{appendix:diff_op_r_manifold}.  A more detailed discussion of continuum mechanics and differential geometry also can be found in~\cite{Marsden1994,Abraham1988}. 

\subsection{Conservation Laws on Manifolds}
\label{sec_conserv_laws}
We discuss briefly how to formulate continuum mechanics in the covariant form on curved surfaces and more generally on manifolds.  There are a few different ways that one can attempt to develop the equations of continuum mechanics in the setting of manifolds.  One approach is to try to use the embedding space with local coordinates and a change of variables.  This can become quite tedious and we shall try to avoid coordinate calculations to the extent it is possible.  We instead use an approach related to the Green-Rivlin-Naghdi Theorem~\cite{GreenRivlin1964} which is based on the use of energy balance and invariance of physical laws under symmetries to derive the conservation equations for mass, momentum, and angular momentum~\cite{Marsden1994,Marsden2007,MarsdenOrtiz2006}.  We remark this approach also provides some insights into why these physical laws manifest in some ways differently in curved manifolds relative to the Euclidean case. 

As a starting point we consider a generalization of the Reynold's transport theorem to the setting of a manifold.  This allows us to express how an integral over a moving parcel of material transforms over time in the Eulerian reference frame.  For a $k$-dimensional domain consider the $k$-form $\bs{\alpha}$.  We have from properties of differential forms and the pull-back that
\begin{eqnarray}
\left. \frac{\partial}{\partial s} \int_{\phi_s(\mathcal{U})} \bs{\alpha} \hspace{0.2cm} \right |_{s = t} = \int_{\phi_{s}(\mathcal{U})} \left . \frac{d}{ds} \left(\phi_{t,s}^{*}\bs{\alpha}\right) \hspace{0.2cm} \right |_{s = t} = \int_{\phi_{t}(\mathcal{U})} {L}_{\mb{v}} \bs{\alpha}.
\end{eqnarray}
The $\phi_{t,s}(\mb{x}) = \phi_s(\phi_t^{-1}(\mb{x}))$ denotes the change in configuration from time $t$ to time $s > t$ as in Section~\ref{sec_exterior_calc}.

In the special case when $\bs{\alpha} = f dv$, where $dv = \omega$ is the volume differential form, we have the Lie derivative is
${L}_{\mb{v}} = \dot{f} + \mbox{div}\left(f \mb{v}\right)$ where $\dot{f} = \partial f/\partial {t} + \mb{v}[f]$ where for short we denote by $\mb{v}[f] = (\mb{d}f)[\mb{v}]$.  We also emphasize that the divergence operator for a vector field on the manifold is now $\mbox{div}(\mb{w}) = w^{b}_{|b}$ where in coordinates the $w^{a}_{|b}$ denotes the covariant derivative component corresponding to derivatives in the direction $\partial_{x^b}$ as in Section~\ref{sec_exterior_calc}.  This gives the scalar transport theorem
\begin{eqnarray}
\label{equ_transport_scalar}
\frac{\partial}{\partial t} \int_{\phi_t(\mathcal{U})} f dv 
= 
\int_{\phi_t(\mathcal{U})} 
\dot{f} + \mbox{div}\left(f \mb{v} \right).
\end{eqnarray}
This can be further specialized in the case of a hypersurface where $\mathcal{U} \subset \mathcal{B}$ and $\dim \mathcal{S} = \dim \mathcal{B} + 1$.  In this case we have for $\phi_t(\mathcal{U}) \subset \mathcal{S}$ the surface-scalar transport theorem 
\begin{eqnarray}
\label{equ_transport_surface}
\frac{\partial}{\partial t} 
\int_{\phi_t(\mathcal{U})}
f dv 
= \int_{\phi_t(\mathcal{U})} 
\left(
\dot{f} + f\left(\overline{\mbox{div}} 
\left(\mb{v}_{\parallel} \right) + \mb{v}_n H \right) \right) dv.
\end{eqnarray}
The $H$ denote the mean-curvature of the manifold describing the configuration of the surface~\cite{Pressley2001}.  The $\mb{v} = \mb{v}_{\parallel} + {v}_n \mb{n}$ where $\mb{v}_n = {v}_n \mb{n}$ gives the velocity component normal to the surface in the ambient space.  Here the $\overline{\mbox{div}}$ denotes the covariant divergence in the manifold describing the surface.  In coordinates aligned with the surface this would be $\overline{\mbox{div}(\mb{w})} = w^{a}_{|a}$ where now the indices $a$ are restricted only over directions $\partial_{x^a}$ corresponding to the components tangent to the surface.   We again emphasize that mechanics requires use of some structure of the ambient physical space in order to make sense of physical quantities such as momentum of the surface~\cite{Marsden1994}.  We shall avoid in our derivation the need to integrate vector fields over the manifold with explicit reference to the ambient space by considering how the mechanics arises from an energy balance principle.  This involves the integration and use of transport theorems only for scalar fields.  We can express the balance of energy for a mechanical system in the general manifold setting as
\begin{eqnarray}
\label{equ_energy_balance}
\frac{\partial}{\partial t} 
\int_{\phi_t(\mathcal{U})}
\rho \left(e + 
\frac{1}{2}
\left\langle \mb{v},\mb{v} \right\rangle   
\right) dv
=
\int_{\phi_t(\mathcal{U})} 
\rho 
\left \langle \bar{\mb{b}},\mb{v} \right\rangle 
dv
+ \int_{\partial \phi_t(\mathcal{U})}
\left \langle \mb{t},\mb{v} \right\rangle.
\end{eqnarray}
The $e$ denotes the energy density per unit mass, $\rho \mb{v}$ the momentum density, $\rho\bar{\mb{b}}$ the body force, and $\mb{t} = \bs{\sigma}[\mb{n}]$ the internal material traction stress vector.  This describes the rate at which the total energy (potential + kinetic) is changing in the system as a consequence of mechanical work done by the body force and stresses.  We point out that the structure of the ambient physical space is still playing a role but now is contained within the inner-products that appear in equation~\ref{equ_energy_balance} given in equation~\ref{equ_manifold_inner_prod}. 

We now use augmentations of the motion $\phi_t$ by diffeomorphisms $\xi_t$ to obtain a new motion $\tilde{\phi}_t = \xi_t \circ \phi_t$.  Since the mechanics associated with any steady translational motion should adhere to Galilean invariance~\cite{LandauMechanics3Ed1976}, we have the new motion $\tilde{\phi}_t$ should satisfy the same energy balance as in equation~\ref{equ_energy_balance}.  We can also consider how the energy principle transforms under other motions such as  steady rotational motion which while non-inertial does preserve distances within the material.  From these considerations, we consider how the energy transforms under the augmented motions $\tilde{\phi}_t$ with diffeomorphisms
\begin{eqnarray}
\label{equ_diffeomorphisms}
\xi_t(\mb{x}) = \mb{x} + (t - t_0)\mb{c} \hspace{0.5cm} \mbox{or} \hspace{0.5cm}
\xi_t(\mb{x}) = \exp\left((t - t_0)\bs{\Omega}\right)\mb{x}.\end{eqnarray}
The $\mb{c}$ is the steady translational velocity and $\bs{\Omega}$ is any anti-symmetric matrix giving the steady angular velocity when we express the rotation as $\mb{R}(t) = \exp\left((t - t_0)\bs{\Omega}\right)$.  

Under the new motion $\tilde{\phi}_t$ in the translational case, we have again that equation~\ref{equ_energy_balance} holds when we substitute in the quantities $\tilde{\phi}_t$, $\tilde{e}(x,t) = e(x,t)$, $\tilde{\rho}(x,t) = \rho(x,t)$, $\tilde{\mb{v}} = \mb{v} + \mb{c}$, $\tilde{\mb{b}} = \mb{b} + \mb{a}$, $\tilde{\mb{t}} = T\xi_t \mb{t}$.  The $T\xi_t$ denotes the tangent map also equivalently the push-forward of $\xi_t$~\cite{Abraham1988}. The $\mb{a} = \dot{\mb{v}} = \partial_t \mb{v} + \nabla_{\mb{v}} \mb{v}$ is the acceleration expressed in Eulerian reference frame with $\nabla_{\mb{v}}$ the covariant derivative of equation~\ref{equ_cov_deriv}.  We use the result of the surface-transport theorem given in equation~\ref{equ_transport_surface} and subtract the original energy balance equation~\ref{equ_energy_balance}.  At time $t = t_0$ we have that $\xi_{t_0} = \mbox{id}$ and the terms of the energy balance remaining from the difference between the motions is
\begin{eqnarray}
\label{equ_net_energy_balance_after_transforms}
\int_{\phi_t(\mathcal{U})} \left(
\dot{\rho} + \rho\left(\overline{\mbox{div}} 
\left(\mb{v}_{\parallel} \right) + \mb{v}_n H \right) \right)
\left(
\langle \mb{v},\mb{c} \rangle + \frac{1}{2}\langle \mb{c},\mb{c} \rangle 
\right) dv = \int_{\phi_t(\mathcal{U})} 
\langle 
\rho(\mb{b} - \mb{a} + \overline{\mbox{div}}(\bs{\sigma}), \mb{c}
\rangle 
dv.
\end{eqnarray}
This must hold for all choices of $\mb{c}$.
To obtain this expression we also used the divergence theorem so that the contributions of the stress vector $\mb{t}$ over the boundary $\partial \phi_t(\mathcal{U})$ can be expressed in terms of integration of the divergence of the stress $\bs{\sigma}$ over the volume.  Since $\mathcal{U}$ is arbitrary and if we take $\mb{c} = q\tilde{\mb{c}}$, the equation~\ref{equ_net_energy_balance_after_transforms} can be localized to a point-wise statement relating the integrands.  We remark localization does require integrands to be sufficiently smooth which we shall assume throughout.  By dividing both sides by $q^2$ and taking the limit $q \rightarrow \infty$ we have that only the quadratic term $\frac{1}{2} \langle \tilde{\mb{c}},\tilde{\mb{c}} \rangle$ persists with all other terms on both sides going to zero.  This requires the multiplying factor of the quadratic term involving $\tilde{\mb{c}}$ to vanish since all the other terms are zero.  This yields the equations for conservation of mass on the surface
\begin{eqnarray}
\label{equ_derive_consv_mass}
\dot{\rho} + \rho\left(\overline{\mbox{div}} 
\left(\mb{v}_{\parallel} \right) + \mb{v}_n H\right).
\end{eqnarray}
We can use this result to further simplify 
equation~\ref{equ_net_energy_balance_after_transforms} to obtain
\begin{eqnarray}
0 = \int_{\phi_t(\mathcal{U})} 
\langle 
\rho(\mb{b} - \mb{a} + \mbox{div}(\bs{\sigma}), \mb{c}
\rangle 
dv.
\end{eqnarray}
Since $\mb{c}$ and $\mathcal{U}$ are arbitrary this gives the equations for conservation of momentum on the surface
\begin{eqnarray}
\label{equ_derive_consv_momentum}
\rho \dot{\mb{v}} = \overline{\mbox{div}}\left(\bs{\sigma}\right) + \rho \mb{b}.
\end{eqnarray}
We use here that $\mb{a} = \rho\dot{\mb{v}} = \rho(\partial_t \mb{v} + \nabla_{\mb{v}} \mb{v})$ gives the acceleration in the Eulerian reference frame with $\nabla_{\mb{v}}$ the covariant derivative of equation~\ref{equ_cov_deriv}.
We can similarly use the transformation under the rotational motions given by the second diffeomorphism in equation~\ref{equ_diffeomorphisms} to get the conservation of angular momentum.  This has the important consequence that the stress tensor $\bs{\sigma}$ must be symmetric in the sense $\sigma^{ab} = \sigma^{ba}$.

These derivations help us to formulate the proper conservation equations of continuum mechanics in the manifold setting.  Similar techniques have also be used to derive other equations useful in elasticity and in constitutive modeling in~\cite{Marsden1994,Marsden2007,Guven2018,MarsdenOrtiz2006}.  We can already see that conservation of mass has some interesting features quite distinct from the flat Euclidean setting, see equation~\ref{equ_derive_consv_mass}.  While the momentum equations appear to look superficially similar to the Euclidean setting, it is important to emphasize that the divergence operator $\overline{\mbox{div}}$ is based on covariant derivatives having in fact quite distinct behaviors from local curvature than the Euclidean setting.  We shall see these results have a number of interesting consequences for constutitive laws, such as the proper form for modeling surface Newtonian fluids for a curved interface.

\subsection{Formulating Hydrodynamic Equations for Curved Fluid Interfaces}

\label{sec:hydro_formulate}

We would like to formulate in a covariant manner the equations of hydrodynamics in the case of a fluid interface that is an incompressible Newtonian fluid.  Using our results from Section~\ref{sec_conserv_laws} we can express the conservation of mass and momentum as
\begin{eqnarray}
\label{equ_cont_mech_fluid}
\left\{
\begin{array}{llll}
\rho \dot{\mb{v}} & = & \overline{\mbox{div}}\left(\bs{\sigma}\right) + \rho\bar{\mb{b}}\\
\dot{\rho} + \rho\left(\overline{\mbox{div}} 
\left(\mb{v}_{\parallel} \right) + \mb{v}_n H\right) & = & 0.
\end{array}
\right.
\end{eqnarray}
We focus here on the the steady-state hydrodynamics of incompressible Newtonian fluids within a curved surface of fixed shape.  This corresponds to $\overline{\mbox{div}} 
\left(\mb{v}_{\parallel}\right) = 0$ and $\mb{v}_n = 0$.  From equation~\ref{equ_cont_mech_fluid}, this yields that $\dot{\rho} = 0$ corresponding to a constant mass density $\rho = \rho_0$ within the interface.  The steady-state hydrodynamics corresponds to the case with $\rho\dot{\mb{v}} = 0$ reducing equation~\ref{equ_cont_mech_fluid} to the conditions 
\begin{eqnarray}
\label{equ_cont_mech_gen_steady}
\left\{
\begin{array}{llll}
\overline{\mbox{div}}\left(\bs{\sigma}\right)
& = & -\mb{b}\\
\overline{\mbox{div}} 
\left(\mb{v} \right) & = & 0.
\end{array}
\right.
\end{eqnarray}
In the notation, we take throughout $\mb{b} = \rho_0 \bar{\mb{b}}$ and $\mb{v} = \mb{v}_\parallel$ since we shall consider only incompressible fluid flows where the velocity is tangent to the surface.  For an incompressible Newtonian fluid the constitutive law should depend on the local rate of deformation $\mb{D}$ of the material as $\bs{\sigma} = \mu_m \mb{D} - p \mathcal{I}$.  The $p$ corresponds to the pressure which acts as a Lagrange multiplier imposing incompressibility.  The $\mu_m$ corresponds to the dynamic shear viscosity of the interfacial fluid.  The $\mathcal{I}$ is the metric associated identity operator.  The rate-of-deformation tensor is $\mb{D} = \frac{1}{2} \partial \mb{C} / \partial{t}$ where $\mb{C}$ is the right-Cauchy-Green Tensor associated with the motion $\phi_t$~\cite{Marsden1994}.  We can express the rate-of-deformation tensor in terms of the covariant derivative as $D_{ab} = v_{a|b} + v_{b|a}$.  For a two-dimensional incompressible fluid having a velocity field that is always tangent to the curved surface, we have $\overline{\mDiv}(\mb{D})^{\flat} =  
-\bs{\delta} \mb{d} \mb{v}^{\flat} + 2K\mb{v}^{\flat}$~\cite{Steigmann1999,AtzbergerSoftMatter2016,ArroyoRelaxationDynamics2009}.  We can further express the divergence in covariant form and exterior calculus operations as  $\overline{\mbox{div}} \left(\mb{v} \right)^{\flat} = -\bs{\delta} \mb{v}^{\flat}$.  From equation~\ref{equ:gradDivCur} we have $\overline{\mDiv}(p \mathcal{I})^{\flat} = \mb{d}p$.  This allows us to express the Stokes hydrodynamic equations in covariant form as
\begin{eqnarray}
\label{equ_Stokes_geometric}
\label{equ_v_stokes}
\left\{
\begin{array}{llll}
\mu_m \left(-\bs{\delta} \mb{d} \mb{v}^{\flat} + 2 K \mb{v}^{\flat} \right) 
- \gamma \mb{v}^{\flat} - \mb{d}p 
& = & - \mb{b}^{\flat} \\
-\bs{\delta} \mb{v}^{\flat} & = & 0.
\end{array}
\right.
\end{eqnarray}
We also added to this equation a phenomenological drag term $-\gamma\mb{v}^{\flat}$ which acts as a force density to model the coupling of the fluid flow on the surface to the bulk three-dimensional surrounding fluid.  This has been done in the context of interfaces such as flat lipid membranes in~\cite{Seki2007}.  Having some form of dissipative tractional stress is important with the surrounding bulk fluid since it provides a model of the physical processes necessary to surppress the otherwise well-known Stokes paradox that arises in purely two-dimensional fluid equations~\cite{Acheson1990,batchelor2000}.  Of course this model is only approximate and one could of course consider more sophisticated hydrodynamic coupling models~\cite{Saffman1976,Lamb1895}.  For specialized cases, such as flat interfaces or spherical geometry, the traction coupling with the bulk fluid flow can be worked out in detail analytically or through asymptotic approximations as done in~\cite{Saffman1976,Lamb1895,AtzbergerSoftMatter2016,
LevineViscoelastic2002,ArroyoRelaxationDynamics2009}.  In the more general setting, solution for the surrounding three-dimensional bulk flow is typically difficult to obtain analytically requiring instead development of separate numerical solvers.  While incorporating such a solver into our approaches is conceptually relatively straight-forward, in practice it involves some significant investments for implementation and handling additional technical issues.  We focus here in this paper on the solver for the surface part of the hydrodynamics.  

\subsubsection{Hodge Decomposition of the Velocity Field and Vector Potentials}
\label{sec_hodge_decomp}
We use a surface Hodge decomposition to derive a formulation of the hydrodynamics that handles the incompressibility constraint for the flow field.  For a fluid within a general manifold we can express the Hodge decomposition using the exterior calculus as
\begin{eqnarray}
\label{equ_hodge_decomp}
\mb{v}^{\flat} = \mb{d}\psi + \bs{\delta}\phi + \mb{h}.
\end{eqnarray}
The $\psi$ is a $0$-form, $\phi$ is a $2$-form, and $\mb{h}$ is a harmonic $1$-form on the surface with respect to the Hodge Laplacian $\Delta_H \mb{h} = \left(\bs{\delta}\mb{d} + \mb{d}\bs{\delta}\right) \mb{h} = 0$.  The dimensionality of the null-space of the Hodge Laplacian depends on the topology of the manifold~\cite{JostBookDiffGeo1991}.  As a consequence, we have for different topologies that the richness of the harmonic differential forms $\mb{h}$ appearing in equation~\ref{equ_hodge_decomp} will vary.  Fortunately, in the case of spherical topology the surface admits only the trivial harmonic $1$-forms $\mb{h} = \bs{0}$ making this manifold relatively easy to deal with in our physical descriptions.  
The incompressibility constraint when applied to equation~\ref{equ_hodge_decomp} results
in $\bs{\delta} \mb{v}^{\flat} = \bs{\delta}\mb{d} \psi = \Delta_H \psi = 0$ which for spherical topology requires $\psi = C$ and $\mb{d}\psi = 0$.  This yields that for an incompressible hydrodynamic flow on the radial manifold surface our physical description must be of the form
\begin{eqnarray}
\label{equ_hodge_decomp2}
\mb{v}^{\flat} = \bs{\delta}\phi + \mb{v}_0^{\flat}.
\end{eqnarray}
Here we have added a velocity $\mb{v}_0$ since this corresponds to the non-tangent contributions from rigid-body translational and rotational motions of the entire interface within the physical ambient space. This can arise physically when the surface force density has a non-zero total net force or torque.  We take throughout the paper the simplification $\mb{v}_0 = 0$ so that the surface velocities should be viewed as accounting for the in-plane contributions of the interface motions.  Of course the flow field in other reference frames can be recovered by adding the non-tangent $\mb{v}_0$ velocity field at each location on the surface.  

Using the Hodge decomposition in equation~\ref{equ_hodge_decomp2}, we see that $\phi$ is a $2$-form on the two-dimensional surface.  We find it convenient to express $\phi$ in terms of a $0$-form using the Hodge star to obtain $\Phi = -\star \phi$.  Using the identities of the Hodge star discussed in Section~\ref{sec_exterior_calc}, we can express the hydrodynamic flow field as 
\begin{eqnarray}
\label{eqn_v_curl_rep}
\mb{v}^{\flat} = -\star\mb{d} \Phi.
\end{eqnarray}
This can be related to classical methods in fluid mechanics by viewing the operator $-\star\mb{d}$ as a type of curl operator that is now generalized to the manifold setting.  The $\Phi$ serves the role of a vector potential for the flow~\cite{Acheson1990,batchelor2000,Lamb1895}.  We substitute equation~\ref{eqn_v_curl_rep} into equation~\ref{equ_v_stokes} and apply the generalized curl operator $\mbox{curl}_\mathcal{M} = -\star\mb{d}$ to both sides to express the fluid equations on the surface as
\begin{eqnarray}
\label{equ_Phi_fluid}
\mu_m\bar{\Delta}^2 \Phi - \gamma \bar{\Delta} \Phi + 2\mu_m (-\star\mb{d}(K(-\star\mb{d})))\Phi &=& \star\mb{d} \mb{b}^{\flat}.
\end{eqnarray}
This provides a particularly convenient form for the fluid equations since it only involves a scalar field on the surface.  We shall utilize primarily this form of the hydrodynamic equations in our numerical methods.

We mention here the importance of distinguishing between the operators when acting on the $0$-forms $\Phi$ in equation~\ref{equ_Phi_fluid} in contrast to the operators acting on $1$-forms $\mb{v}^{\flat}$ in equation~\ref{equ_v_stokes}.  In our notation here we use $\bar{\Delta} = -\bs{\delta}_1\mb{d}_0$ to obtain our surface Laplacian.  The sign convention ensures our surface Laplacian is a negative semidefinite operator.  This provides consistency with intuition that is often used in physical setting and agreement with the standard Laplacian of vector calculus encountered in the Euclidean case.  This is in contrast with the Hodge-Laplacian $\Delta_H = \bs{\delta}\mb{d} + \mb{d}\bs{\delta}$ used in differential geometry which is positive semi-definite~\cite{JostBookDiffGeo1991,Abraham1988}.  Our sign conventions also ensure that our operator $\bar{\Delta}$ is in agreement with the Laplace-Beltrami operator on the surface.  Further distinctions can also arise when interpreting $\mb{d}$ and $\bs{\delta}$ depending on the dimensionality of the manifold and the order $k$ of the $k$-forms on which the operator acts~\cite{Abraham1988}.  We mention these distinctions since in our experience in practice these differences and the sign conventions can become a significant source of book-keeping and confusion in modeling and in numerical methods.  We have found it convenient when formulating our numerical methods and performing implementations to use the Hodge-Laplacian $\Delta_H = -\bar{\Delta} = \bs{\delta}_1\mb{d}_0$ to avoid carrying around the sign.  However, we have found it convenient in our physical analysis and discussions to use the negative semi-definite surface Laplacian $\bar{\Delta}$ as we have discussed for equation~\ref{equ_Phi_fluid}.  Given these considerations, we shall primarily use $\bar{\Delta}$ throughout most of our discussions in this paper unless otherwise noted.

\section{Numerical Methods}
\label{sec_num_methods}

\subsection{Hyperinterpolation for $L^2$-Inner-Products on Manifolds using Lebedev Quadratures}
\label{sec_hyperinterpolation}

We develop spectral methods based on a Galerkin-style approximation by introducing an $L^2$-inner-product $\langle , \rangle_\mathcal{M}$ on the manifold surface $\mathcal{M}$.  To approximate the inner-product with a high order of accuracy we use hyperinterpolation~\cite{Sloan2001} on the manifold.  For the case when the manifold is a sphere we use the Lebedev quadrature~\cite{Lebedev1976,Lebedev1999}.  We introduce here ways to develop high order quadratures for integration on more general manifold surfaces of radial shape.  Our approach is based on the use of the Radon-Nikodym Theorem~\cite{Lieb2001} to relate in a coordinate-free manner the measure associated with surface area on the sphere to the radial manifold.     

\subsubsection{Surface Quadrature for Radial Manifolds}
\label{sec_surf_quad}

\begin{figure}[H]
\centering
\includegraphics[width=0.85\linewidth]{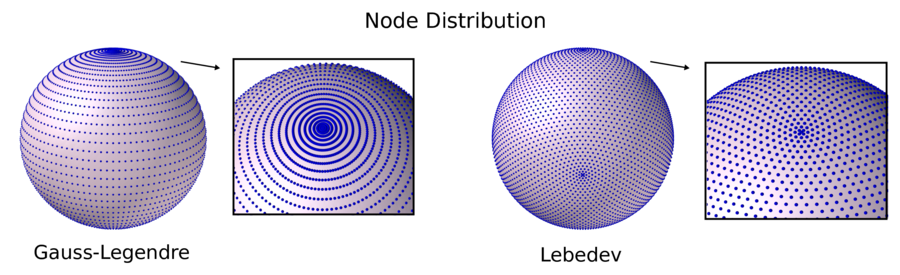}
\caption{Node Distribution of the Gauss-Legendre Quadrature Compared to  Lebedev Quadrature.  We consider the Gauss-Legendre quadrature with 5886 nodes with a comparable Lebedev quadrature having 5810 nodes ($113^{th}$   order of accuracy)~\cite{Lebedev1999}.  We see that the Gauss-Legendre quadrature has dense clustering of points along latitudinal rings when approaching the north and south poles.  We see that while the Lebedev quadrature has some clustering around a few points, these are less dense and overall exhibits sampling with a greater level of symmetry over the sphere. }
\label{fig:quad_nodes_compare}
\end{figure}

We develop our spectral methods based on a Galerkin-style approximation by introducing a metric associated $L^2$ inner-product $\langle , \rangle_\mathcal{M}$ on the manifold surface $\mathcal{M}$.  For any two differential $k$-forms $\bs{\alpha}$ and $\bs{\beta}$ we introduce the manifold inner-product
\begin{equation}
\label{equ_innerProd_M}
\left \langle \bs{\alpha}, \bs{\beta} \right \rangle_\mathcal{M} = \int_\mathcal{M} \left \langle \bs{\alpha}, \bs{\beta} \right \rangle_g dA
\end{equation}
where $\left \langle \bs{\alpha}, \bs{\beta} \right \rangle_g$ is the local metric inner-product on $k$-forms on the manifold.  To compute in practice this inner-product to a high order of accuracy we need to integrate over the manifold.  For this purpose we introduce an approach based on Lebedev quadratures.  The Lebedev quadrature nodes are derived by solving a non-linear system of equations that impose both exactness of integration on spherical harmonics up to a specified order while maintaining symmetry under octahedral rotations and reflections~\cite{Lebedev1976,Lebedev1999}.  One could also consider using a quadrature based on spherical coordinates and sampling on the latitudinal and longitudinal points which have some computational advantages by using the Fast Fourier Transform~\cite{HealyDriscoll1994,Kunis2003,Healy2003}.  However, these nodes have significant asymmetries with nodes forming dense clusters near the poles of the sphere, see Figure~\ref{fig:quad_nodes_compare}.
We favor Lebedev quadratures which while having some localized clustering at a few points exhibits overall a greater level of symmetry and less severe clustering.  Quadrature on the surface of the sphere is still an active area of research with many recent results in the literature investigating the advantages and draw-backs of different methods depending on the intended use and application~\cite{Sloan2001,Hesse2010,Atkinson1982,BeentjesSphereQuadCompare2015}.  We briefly mention that this includes recently introduced Spherical t-Designs~\cite{WomersleySphericalDesigns2017} and nodes obtained by minimizing different types of energies motivated by generalizing the classical Thomson problem~\cite{Thomson1904,BeentjesSphereQuadCompare2015}.  
The recently introduced Spherical t-Designs are also attractive given their overall symmetry of quadrature nodes~\cite{WomersleySphericalDesigns2017}.  However, the Spherical t-Designs only have $60\%$ approximation efficiency in the number of nodes per the accuracy achieved. In contrast the Lebedev quadratures achieve optimal approximation efficiency~\cite{BeentjesSphereQuadCompare2015}.  In principle, almost any quadrature on the sphere could be used within the overall numerical approaches we present.  We use the Lebedev quadratures given their high level of symmetry and approximation efficiency.

\begin{figure}[H]
\centering
\includegraphics[width=0.9\linewidth]{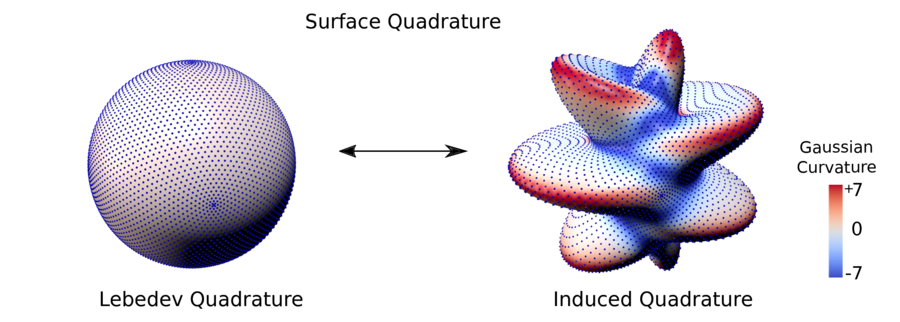}
\caption{Lebedev Quadrature with 5810 nodes.  We show on the left the Lebedev quadratures for integration of functions on a sphere of unit radius.  The Lebedev quadrature integrates exactly all spherical harmonics up to the $131^{\mbox{st}}$ order~\cite{Lebedev1999}.  The mapping of the sphere to the manifold on the right induces a new quadrature weighted by the local manifold metric.  While the induced quadrature is no longer exact for spherical harmonics on the surface it still exhibits a high level of accuracy.  We show by the colors the Gaussian curvature $K$ on the surfaces over the range $-7$ to $7$.}
\label{fig:shapes}
\end{figure}

We obtain a quadrature formula having a high level of accuracy for integration on the radial manifold surface using the manifold metric.  We derive surface quadratures by using a diffeomorphism $\Phi$ that transforms the reference sphere to the radial manifold.  This is done by first considering functions $f$ expressible as a finite combination of spherical harmonics up to the order of accuracy of the quadrature.  These band-limited functions $f$ are integrated exactly by the Lebedev quadrature with nodes $\mb{x}_i$ and weights $w_i$ which we can express using spherical coordinates as  
\begin{eqnarray}
\int_{\mathcal{S}^2} f dA = \int_0^{2\pi}\int_0^{\pi} f \sin(\phi) d\phi d\theta = \sum_i f_i\cdot w_i.
\end{eqnarray}\\
The integration on the manifold surface $\mathcal{M}$ can be expressed using spherical coordinates as
\begin{eqnarray}
\label{eqn_quad_manifold_coord}
\int_\mathcal{M} f dA = \int_0^{2\pi}\int_0^{\pi} f \sqrt{|g|} d\phi d\theta 
= \int_0^{2\pi}\int_0^{\pi} f \frac{\sqrt{|g|}}{\sin(\phi)}\sin(\phi) d\phi d\theta 
= \int_{\mathcal{S}^2} f \frac{\sqrt{|g|}}{\sin(\phi)} dA.
\end{eqnarray}
The term $\sin(\phi)$ vanishes at the poles and must be considered carefully.  Since the metric has been expressed relative to spherical coordinates, we have for radial manifolds generated by diffeomorphisms that the $\sqrt{|g|}$ vanishes at the pole and even more importantly the ratio $\sqrt{|g|}/\sin(\phi)$ approaches a finite value in the limit of approaching a pole.  To derive a quadrature on the surface it is useful to give an alternative view on our derivation of equation~\ref{eqn_quad_manifold_coord} more abstractly without relying on coordinates.  We can pull-back integration to be on the reference sphere and consider the change of measure for areas that must occur when transforming from the unit sphere to the radial manifold.  The pull-back of the radial manifold area measure $\mu_{\mathcal{M}}$ to the sphere is always absolutely continuous with respect to the sphere area measure $\mu_{\mathcal{S}^2}$.  By the Radon-Nikodym Theorem~\cite{Lieb2001}, this allows us to express without reference to coordinates the relationship between the integrations as 
\begin{eqnarray}
\int_{\mathcal{M}} f d\mu_{\mathcal{M}} = \int_{\mathcal{S}^2} f \eta d\mu_{\mathcal{S}^2}.
\end{eqnarray}
The $\eta(\mb{x}) = d\mu_{\mathcal{M}} / \mu_{\mathcal{S}^2}$ denotes the Radon-Nikodym derivative at $\mb{x}$~\cite{Lieb2001}.  We make the correspondence  
$dA = d\mu_{\mathcal{M}}$ with $dA$ to be understood by context as the area element on the manifold $\mathcal{M}$.  We also make the correspondence $dA = d\mu_{\mathcal{S}^2}$ to be understood by context as the area element on the sphere surface.  From these considerations, we have for any chart of spherical coordinates $(\bar{\theta},\bar{\phi})$ that
\begin{eqnarray}
\int_{\mathcal{M}} f dA = \int_{\mathcal{S}^2} f \eta dA = \int_{\mathcal{S}^2} f \frac{\sqrt{|\bar{g}|}}{\sin(\bar{\phi})} dA.
\end{eqnarray}
The first two expressions are coordinate-free whereas the last expression depends on the chosen spherical coordinates $(\bar{\theta},\bar{\phi})$.  Since this correspondence holds for integration over any smooth subset of the manifold, we have that the Radon-Nikodym derivative at $\mb{x}$ can be expressed for any two choices of spherical coordinates $(\theta,\phi)$ and $(\tilde{\theta},\tilde{\phi})$ as
\begin{eqnarray}
\eta(\mb{x}) = \frac{\sqrt{|{g}|}}{\sin({\phi})} = \frac{\sqrt{|\tilde{g}|}}{\sin(\tilde{\phi})}.
\end{eqnarray}
This shows that the ratio that arises does not depend on the particular choice of coordinates.  This is useful in numerical calculations since we can use coordinate charts so that any location $\mb{x}$ on the sphere the ratio has a non-vanishing denominator.  We use primarily two coordinate charts throughout our calculations.  The first with the poles along the $z$-axis (north and south poles) and the second with the poles along the $x$-axis (east and west poles)~\cite{AtzbergerSoftMatter2016}.  We denote these by $(\theta,\phi)$ and $(\tilde{\theta},\tilde{\phi})$.  For instance, when the denominator would be too close to zero for reliable numerical calculation in the first chart, we switch to the second chart and compute ${\sqrt{|\tilde{g}|}}/{\sin(\phi)}$ where $\tilde{g}$ is the metric expressed in the other chart coordinates. 

These results give us a way to use a quadrature on the sphere to induce a quadrature on the manifold surface as
\begin{eqnarray}
\label{equ_surf_quad}
\int_{\mathcal{M}} f dA = \sum_i f_i \cdot \bar{w}_i.
\end{eqnarray}
The induced weights are given by $\bar{w}_i = \eta(\mb{x}_i) w_i = {\sqrt{|g|}}/{\sin(\phi)} w_i$ and nodes $\mb{z}_i = \Phi(\mb{x}_i)$.  In the case of Lebedev quadratures and radial manifolds the quadrature is no longer exact for spherical harmonics.  Instead the quadrature is exact for the collection of functions of the form $f = Y/\eta$ where $Y$ is a finite combination of spherical harmonics.  In practice, we find that the induced quadrature still exhibits a high level of accuracy for smooth fields on the radial manifolds we consider.  We use these results to compute the manifold $L^2$-inner-product for the surface by $\left \langle \bs{\alpha}, \bs{\beta} \right \rangle_{\mathcal{M}} \approx \left \langle \bs{\alpha}, \bs{\beta} \right \rangle_Q$ with
\begin{eqnarray}
\label{equ_innerProd_M_quad}
\left \langle \bs{\alpha}, \bs{\beta} \right \rangle_Q = \sum_i \left \langle \bs{\alpha}(\mb{x}_i), \bs{\beta}(\mb{x}_i) \right \rangle_g \cdot \bar{w}_i.
\end{eqnarray}

\subsubsection{Validation of Surface Quadrature using Gauss-Bonnet Theorem}
\label{sec_quad_validation}

We show the efficacy of this approach for radial manifolds.  Since in general it is not straight-forward to obtained closed-form analytic solutions for surface integrals on general radial manifolds, we develop a test based on the Gauss-Bonnet Theorem~\cite{Pressley2001,SpivakDiffGeo1999}. Since each of the
 manifolds have spherical topology, a consequence of the Gauss-Bonnet Theorem is that the Gaussian curvature when integrated over the surface must have
\begin{eqnarray}
\int_{\mathcal{M}} K(\mb{x}) dA = 2\pi \chi(\mathcal{M}).
\end{eqnarray}
The $\chi(\mathcal{M})$ is the Euler Characteristic of the surface~\cite{Pressley2001,SpivakDiffGeo1999}.  For spherical topology the Euler Characteristic is $\chi(\mathcal{M}) = 2$ requiring for all the radial manifolds that $\int_{\mathcal{M}} K(\mb{x}) dA = 4\pi$.  We perform this calculation using our surface quadrature introduced in equation~\ref{equ_surf_quad}.  

\begin{figure}[H]
\centering
\includegraphics[width=\linewidth]{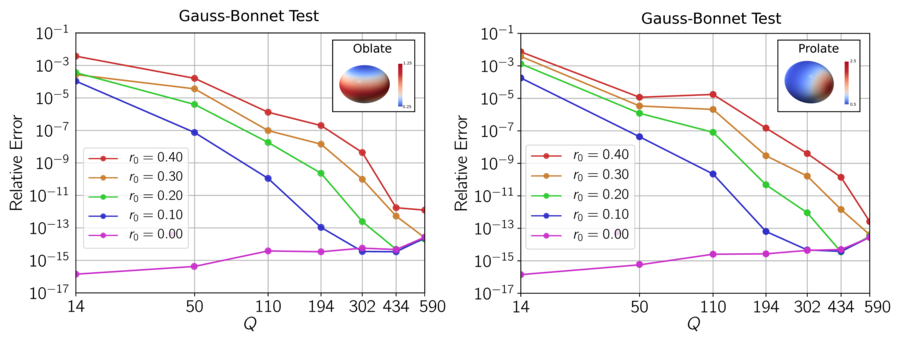} 
\caption{Quadrature on Radial Manifolds.  For ellipsoids of oblate and prolate shapes we test the quadrature by integrating the Gaussian curvature over the manifold and compare the results with the predictions of the Gauss-Bonnet Theorem~\cite{SpivakDiffGeo1999,Pressley2001}.  We show the accuracy of the quadrature as the number of quadrature nodes $Q$ increases and when varying the shape parameter $r_0$ of the ellipsoid, see equation~\ref{equ_ellipsoid_shape_oblate} and equation~\ref{equ_ellipsoid_shape_prolate}.  The case $r_0 = 0$ gives a sphere with the other $r_0$ values giving the shapes as shown in Figure~\ref{fig:ellipsoid_hydro}.  We show as insets the ellipsoids with $r_0 = 0.4$ and the Gaussian curvature distribution on the surface.
}
\label{fig:quadrature_gauss_bonnet_test1}
\end{figure}

This provides a significant test of a number of components of our calculation.  To obtain the correct final result requires that the Gaussian curvature, metric factor, and first and second fundamental forms computed in our calculations properly combine with the Lebedev quadrature to yield the final integral value of $4\pi$.  We show the results of this test for both the oblate and prolate ellipsoidal manifolds as we vary both $r_0$ and the order of the quadrature in Figure~\ref{fig:quadrature_gauss_bonnet_test1}. 

We consider oblate and prolate ellipsoids that can be characterized by a parameter $r_0$ which controls the shape.  In coordinates $(x,y,z)$, the prolate ellipsoid is 'stretched' along the $z$-axis while the oblate is stretched equally along both the $x$ and $y$ directions.  The oblate ellipsoid corresponds to ${(x^2 + y^2)}/{(1+r_0)^2} + z^2 = 1$ with
\begin{eqnarray}
\label{equ_ellipsoid_shape_oblate}
r(\theta,\phi) & = & \frac{1+r_0}{\sqrt{(1+r_0)^2\sin^2(\phi)+ \cos^2(\phi)}} \\  
K(\theta,\phi) & = & \frac{1}{\left(1 + ((1+ r_0)^2 -1 )\cdot \frac{(1+r_0)^2 \cos^2(\phi)}{(1+r_0)^2 \cos^2(\phi) + \sin^2(\phi)}\right)^2}.
\end{eqnarray}
The prolate ellipsoid corresponds to $x^2 + y^z + {z^2}/{(1+ r_0)^2} = 1$ with
\begin{eqnarray}
\label{equ_ellipsoid_shape_prolate}
r(\theta,\phi) &=& \frac{1 + r_0}{\sqrt{1 + (((1+ r_0)^2 - 1)\sin^2(\phi)}} \\
K(\theta,\phi) &=& \frac{(1+r_0)^6}{\left((1+r_0)^4 + (1-(1+r_0)^2)\cdot\frac{(1+r_0)^2 \cos^2(\phi)}{(1+r_0)^2\sin^2(\phi)+ \cos^2(\phi)}\right)^2}. 
\end{eqnarray}
The $K$ denotes the Gaussian curvature and the $r$ is the shape function of the radial manifold as discussed in Appendix~\ref{appendix:radialManifold}.  
We vary $r_0$ to obtain different ellipsoidal shapes as shown in Figure~\ref{fig:ellipsoid_hydro}.

For the oblate and prolate ellipsoids, we see that in each case as the number of quadrature nodes increases the computed approximation to the integral converges rapidly to the Euler characteristic of the surface $2\pi \chi(\mathcal{M}) = 4\pi$.  We further see that even on the non-spherical manifolds the rate of convergence is super-algebraic as the number of quadrature points increase.  This indicates that despite the distortions and re-weighting induced by the transformation of Lebedev nodes the radial manifold quadrature still retains a high level of accuracy, see Figure~\ref{fig:quadrature_gauss_bonnet_test1}.

\begin{figure}[H]
\centering
\includegraphics[width=\linewidth]{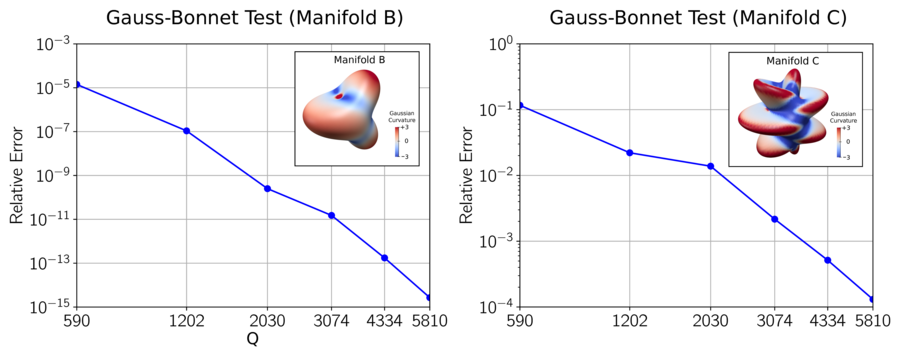} 
\caption{Quadrature on Radial Manifolds.  We test our quadratures by integrating the Gaussian curvature on the manifold and comparing with the predictions of the Gauss-Bonnet Theorem~\cite{SpivakDiffGeo1999,Pressley2001}.  We show the relative errors as the number of quadrature points $Q$ increases in the case of $r_0 = 0.3$ for the manifolds B and C given by equation~\ref{equ_manifold_B}.  Since the Gaussian curvature is not known analytically in advance for these manifolds the test also validates the geometric approximations made in our numerical methods.
}
\label{fig:quadrature_gauss_bonnet_test2}
\end{figure}

We perform further investigation of the quadrature methods on two radial manifolds having the more complicated shapes of Manifold B and Manifold C shown in Figure~\ref{fig:quadrature_gauss_bonnet_test2} and Figure~\ref{fig:manifold_shapes}.  For these shapes, the Gaussian curvature on the surface is not known analytically in advance.  These tests provide a stronger test and validation than just the quadratures since contributions from errors arise also from the geometric approximations made by our numerical methods discussed in Section~\ref{sec_num_methods} and Appendix~\ref{appendix:radialManifold}.  Our results for these tests are shown in Figure~~\ref{fig:quadrature_gauss_bonnet_test2}.

We find that while the combined sources of approximation yield larger errors we can still obtain overall small errors with a sufficient number of quadrature nodes.  The Manifold C shape has a few regions of especially large Gaussian curvatures which provides a useful challenge for the numerical methods.  We find that once the spherical harmonic basis captures sufficiently these features of the geometry the quadrature then converges rapidly.  In summary, we find that given a sufficient number of quadrature nodes we can use our extended Lebedev quadratures of equation~\ref{equ_surf_quad} to obtain accurate integration on radial manifolds.

\subsection{Galerkin Approximation of Partial Differential Equations on Manifolds}
\label{sec_Galerkin_approx}
We develop spectral numerical methods based on Galerkin approximation~\cite{BrennerFEM2008}.  Our approach uses hyperinterpolation of functions for $L^2$-projection to spherical harmonics based on Lebedev quadrature~\cite{Lebedev1999,Sloan2000}.  We consider for radial manifolds the partial differential equations of the form
\begin{eqnarray}
\label{equ_main_PDE}
\mathcal{L} \mb{u} = \mb{\mathfrak{g}}.
\end{eqnarray}
We consider $\mathcal{L}$ that are linear operators that take $k$-forms to $m$-forms.  We take $\mb{\mathfrak{g}}$ to be a general $m$-form not to be confused with the metric tensor $\mathbf{g}$ discussed in Section~\ref{sec_exterior_calc}.  Typical differential operators $\mathcal{L}$ encountered in practice include operators that arise from composition of the exterior derivative $\mb{d}$ and Hodge star $\star$ as discussed in Section~\ref{sec_cont_mechanics}.  This includes the Laplace-Beltrami operator $\Delta_{LB} = -\bs{\delta}_1\mb{d}_0$ which takes $0$-forms to $0$-forms and the Hodge Laplacian $\Delta_H = \bs{\delta}_2 \mb{d}_1 + \mb{d}_0 \bs{\delta}_1$ which takes $1$-forms to $1$-forms.  The subscripts here indicate the order of differential form upon which the operators act.  

To obtain numeric methods for equation~\ref{equ_main_PDE}, we consider Galerkin approximations based on the weak formulation 
\begin{eqnarray}
\label{equ_weak_form}
\left\langle \mathcal{L} \mb{u}, \bs{\psi} \right\rangle  = \left\langle \mb{\mathfrak{g}}, \bs{\psi} \right\rangle.
\end{eqnarray}
Here the $\bs{\psi}$ are test differential $m$-forms. We take the inner-product $\left\langle  \cdot \right \rangle = \left\langle  \cdot \right \rangle_{\mathcal{M}}$ to be the manifold metric associated inner-product on $m$-forms defined in equation~\ref{equ_manifold_inner_prod}.  We denote the corresponding Hilbert space of square integrable differential $m$-forms as $\Lambda^2_m(\mathcal{M})$~\cite{Lieb2001,Bochev2006}.  Central to our approximation is a choice of the finite dimensional subspace of $\Lambda^2_m(\mathcal{M})$.  We use for our solution space the finite subspace of differential forms $\psi$ that are dual to finite spherical harmonic expansions.  In particular, we consider test $m$-forms that correspond to the surface viewed as a submanifold of the ambient space.  This allows in numerical calculations a way to more readily obtain global test $m$-forms $\bs{\psi}$.  We take for given coordinates of the ambient space the test $m$-forms to be those forms that can be expressed using a finite spherical harmonics expansion of the form $\bs{\psi} = \psi_{i_1\ldots i_k} \mb{d}x^{i_1} \wedge \ldots \wedge \mb{d}x^{i_k}$ with each component having finite expansion $\psi_{i_1\ldots i_k} = \sum\hat{\psi}^{(\ell)}_{i_k} Y_{\ell}$ where $i_k \in \{1,2,3\}$ and $Y_{\ell}$ is a spherical harmonic mode as in Appendix~\ref{appendix:sphericalHarmonics}.  We expand using spherical harmonics up to order $\lfloor L/2 \rfloor$ since we take our quadratures up to order $L$ as discussed in Section~\ref{sec_surf_quad}.  We remark that this general approach reduces to a finite spherical harmonics expansions of functions in the case of $0$-forms.

We compute numerical approximations of the inner-products on the manifold surface using our surface-induced quadratures developed in Section~\ref{sec_surf_quad} to obtain 
\begin{eqnarray}
\left \langle \mathcal{L} \bar{u}, \bs{\psi} \right \rangle_Q  = \left \langle \bar{\mathfrak{g}}, \bs{\psi} \right \rangle_Q.
\end{eqnarray}
When the manifold is a sphere the quadrature exactly computes the manifold inner-product when taken up to order $L$.  In the case of the more general radial manifolds the quadrature introduces some additional source of errors in the calculation.  Also in practice the operators $\mathcal{L}$ considered often have significant dependence on geometric features of the manifold.  We use throughout an isogeometric approach with the manifold shape represented as a finite spherical harmonics expansion of order $L$.  For radial manifolds this corresponds to an expansion of the function $r(\theta,\phi)$ and then computing related geometric quantities using the derivatives of the spherical harmonics as discussed in Appendix~\ref{appendix:radialManifold} and Appendix~\ref{appendix:sphericalHarmonics}.  Given these approximations we obtain the final linear system of equations for $\mb{\hat{u}}$ 
\begin{eqnarray}
\label{equ_Lu_Mg}
L \hat{\mb{u}} & = & M \hat{\mb{\mathfrak{g}}}.
\end{eqnarray}
The $\mb{\hat{u}}$ and $\mb{\hat{\mathfrak{g}}}$ denote the collection of coefficients in the expansions of $\bar{u}$ and $\bar{\mathfrak{g}}$.  The $L$ denotes the stiffness matrix and $M$ denotes the mass matrix~\cite{BrennerFEM2008}.

We mention that this weak-form associated with the manifold metric offers some particular conveniences when $\mathcal{L}$ is a differential operator that can be written as a composition of exterior calculus operators.  This allows for natural use to be made of the adjoint operators to lower the differential order of the equations as is typically done in the Euclidean setting by use of integration by parts.  For instance, for the Laplace-Beltrami operator we have $\left\langle \bs{\delta}\mb{d} u, \psi\right\rangle = \left\langle \mb{d} u, \mb{d}\psi\right\rangle$ by using the adjoint property of $\mb{d}$ and $\bs{\delta}$ given in equation~\ref{equ_adjoint_d_delta}.  A similar approach can be carried out for other operators.

\subsubsection{Approximating the Hydrodynamic Equations on Radial Manifolds}
We develop numerical methods using Galerkin approximation for the  hydrodynamic equations we formulated in Section~\ref{sec:hydro_formulate} for curved fluid interfaces.  We consider how to approximate equation~\ref{equ_Phi_fluid} using the corresponding weak formulation.  We then discuss some details of how we handle the different terms.  We first give general expressions for the terms in equation~\ref{equ_Lu_Mg} and then discuss how these expressions are approximated numerically.

We showed that for incompressible fluids on a surface governed by equation~\ref{equ_Stokes_geometric} the Hodge decomposition of equation~\ref{equ_hodge_decomp} could be used to obtain equation~\ref{equ_Phi_fluid}.  This allows us to express the fluid velocity in terms of a vector potential $\Phi$.  For a force density $\mb{b}$ driving the fluid and using the generalized curl, we have $\Xi = \star\mb{d} \mb{b}^{\flat} = -\mbox{curl}_{\mathcal{M}}(\mb{b}^{\flat})$ which gives the RHS term of equation~\ref{equ_Phi_fluid}.  We represent these fields on the radial manifold numerically using truncations of the spherical harmonics expansions
\begin{eqnarray}
\Phi = \sum_{\ell} \hat{\Phi} Y_{\ell}, \hspace{0.5cm} \Xi = \sum_{\ell} \hat{\Xi} Y_{\ell}.
\end{eqnarray}
We use the orthogonality and normalization of the spherical harmonics discussed in Appendix~\ref{appendix:sphericalHarmonics}.  We compute the differential operators using the expressions in Appendix~\ref{appendix:radialManifold} and~\ref{appendix:diff_op_r_manifold}.  In the case we are given the force density $\mb{b}^{\flat}$ we can construct the term $M \mb{\mathfrak{\hat{g}}}$ in equation~\ref{equ_Lu_Mg} by computing
\begin{eqnarray}
[M \mb{\hat{\mathfrak{g}}}]_i = \langle \Xi, Y_i \rangle_\mathcal{M} = -\langle -\star \mb{d} \mb{b}^{\flat}, Y_i \rangle_\mathcal{M}.
\end{eqnarray}
In the case when we instead are given $\Xi$ or the expansion coefficients of $\hat{\Xi}$, we alternatively compute the product $M \mb{\mathfrak{\hat{g}}}$ from
\begin{eqnarray}
[\mb{\hat{\mathfrak{g}}}]_i & = & \hat{\Xi} \\
\lbrack M \rbrack_{ij} & = & \langle Y_j, Y_i \rangle_\mathcal{M}.
\end{eqnarray}
The stiffness tensor $L$ for the Stokes equations can be expressed in terms of our surface Laplacian $\bar{\Delta}$ in equation~\ref{equ_Phi_fluid} as
\begin{eqnarray}
L_{ij} = \left\langle \left(\mu_m\bar{\Delta}^2 - \gamma \bar{\Delta} + 2\mu_m \left(-\star\mb{d}(K(-\star\mb{d}))\right) \right) Y_i, Y_j \right\rangle_M = \tilde{A}_{ij} + \tilde{B}_{ij} + \tilde{C}_{ij}.
\end{eqnarray}
In this notation, we have $\bar{\Delta} = -\bs{\delta}\mb{d}$, $\tilde{A}_{ij} = \mu_m\left \langle \bar{\Delta}^2 Y_i , Y_j \right\rangle_\mathcal{M}$, $\tilde{B}_{ij} = -\gamma\left\langle \bar{\Delta} Y_i, Y_j \right \rangle_\mathcal{M}$, $\tilde{C}_{ij} = 2\mu_m \left \langle \left(\star\mb{d}(K(\star\mb{d}))\right) Y_i, Y_j\right \rangle$.  In our numerical calculations, we prefer to avoid carrying around the sign and use the equivalent formulation in terms of the Hodge-Laplacian $\Delta_H = \bs{\delta}\mb{d} = -\bar{\Delta}$ as 
\begin{eqnarray}
L_{ij} = \left\langle \left(\mu_m\Delta_H^2 + \gamma \Delta_H  + 2\mu_m \left(-\star\mb{d}(K(-\star\mb{d}))\right) \right) Y_i, Y_j \right\rangle_{\mathcal{M}} = A_{ij} + B_{ij} + C_{ij}.
\end{eqnarray}
In this notation, we have $A_{ij} = \mu_m\left \langle \Delta_H^2 Y_i , Y_j \right\rangle_\mathcal{M}$, $B_{ij} = \gamma\left\langle \Delta_H Y_i, Y_j \right \rangle_\mathcal{M}$,  $C_{ij} = 2\mu_m \left \langle \left(\star\mb{d}(K(\star\mb{d}))\right) Y_i, Y_j\right \rangle$.

As discussed in Section~\ref{sec_exterior_calc} the exterior calculus has the convenient property of allowing us to identify readily adjoint operators and perform calculations in a manner similar to integration by parts done in the Euclidean setting.  We make use of the adjoint relationship between the exterior derivative operator $\mb{d}$ and co-differential operator $\bs{\delta}$ which gives $\left \langle \bs{\delta}u,v \right \rangle = \left \langle u, \mb{d} v \right \rangle$.  
Using these adjoint properties allows us to express the stiffness tensor as

\begin{eqnarray}
\label{equ_FEM_Lij}
L_{ij} & = & A_{ij} + B_{ij} + C_{ij} \\
\label{equ_FEM_Aij}
A_{ij} & = & \mu_m\left\langle \Delta_H Y_j, \Delta_H Y_i \right \rangle_{\mathcal{M}}  \\
B_{ij} & = & \gamma \left\langle \mb{d} Y_j, \mb{d} Y_i \right \rangle_{\mathcal{M}} \\
\label{equ_FEM_Cij}
C_{ij} & = & -2\mu_m \left \langle K (-\star\mb{d}) Y_i, (-\star\mb{d}) Y_j \right\rangle_{\mathcal{M}}.
\end{eqnarray}  
This formulation for the stiffness matrix $L$ is similar to expressions obtained in the Euclidean setting by use of integration by parts~\cite{BrennerFEM2008}.  An advantage of this weak form for the stiffness matrix is that the order of differentiation has now been reduced to order two.  This weak form allows us in equation~\ref{equ_FEM_Aij} to replace evaluation of the bi-harmonic operator $\Delta_H^2$ in equation~\ref{equ_Phi_fluid} with the computation of the Hodge-Laplacian operator $\Delta_{H}$ two times without composition.  The exterior calculus adjoint relationships are particularly useful for $C_{ij}$ where we see that we can avoid the need to take a derivative of the Gaussian curvature $K$ on the surface, see equation~\ref{equ_FEM_Cij}.

We use these expressions for the stiffness matrix to obtain $\tilde{L}$ by approximating each of the manifold inner-products $\langle \cdot , \cdot \rangle_M$ by $\langle \cdot, \cdot \rangle_Q$ using the extended Lebedev quadrature approach we discussed in Section~\ref{sec_surf_quad}.
Computing the stiffness tensor components $\tilde{L}_{ij}$ then is reduced to computing accurate local approximations of the Hodge-Laplacian operator $\Delta_H = \bs{\delta}\mb{d}$, exterior derivative operator $\mb{d}$, and the generalized curl operator $\mbox{curl}_{\mathcal{M}} = -\star\mb{d}$.  This is done by using our spherical harmonics representation of the surface and applying in real-space the evaluation of these operators at the quadrature points to evaluate the needed inner-products.  Expressions for each of these operations is given in Appendix~\ref{appendix:radialManifold} and~\ref{appendix:diff_op_r_manifold}.  In this manner we obtain the terms of equations~\ref{equ_FEM_Lij}--~\ref{equ_FEM_Cij} needed to formulate equation~\ref{equ_Lu_Mg}.  This allows us to approximate numerically solutions of equation~\ref{equ_Phi_fluid} which govern hydrodynamic flows on the surface.

\section{Convergence of the Numerical Solver for Surface Hydrodynamics}
\label{sec_convergence}

We investigate the convergence and accuracy of our hydrodynamics solver.  
For manifolds there are few analytically known solutions against which we can compare the results of our solver.  To address this issue, we construct reference solutions on the manifold using the method of manufactured solutions~\cite{Shih1985,SandiaManufacturedSol2000}.  In the manifold setting we have additional challenges since even computing the action of the differential operators involves non-trivial dependencies on the geometry of the surface.  We show how to handle these issues in the case of ellipsoids having prolate and oblate shapes to obtain reference solutions with high precision against which we can compare the results of our hydrodynamics solver.

We use the method of manufactured solutions in the hydrodynamics setting by specifying a velocity potential $\bar{\Phi}$.  We make a choice for the right-hand side (RHS) force density term $\star \mb{d} \mb{b}^{\flat}$ so that in equation~\ref{equ_Phi_fluid} we would obtain as the solution our specified $\bar{\Phi}$.  To obtain the RHS data with high precision on ellipsoids we evaluate the differential operators on the left-hand-side (LHS) of equation~\ref{equ_Phi_fluid} using symbolic computations~\cite{Sympy2017}.  We then use the RHS data for our numerical methods to solve the hydrodynamic equations~\ref{equ_Phi_fluid} and compare our numerical results $\tilde{\Phi}$ and $\mb{\tilde{v}}$ with the known solutions $\bar{\Phi}$ and $\bar{\mb{v}}^{\sharp} = \left(-\star \mb{d}\bar{\Phi}\right)^{\sharp}$.  

Ellipsoids provide good test manifolds for our methods since they have a level of geometric richness, such as heterogeneous Gaussian curvature, while remaining tractable for symbolic computations.  Also for ellipsoids, we have explicit expressions for many of the intermediate terms, such as the Gaussian curvature, which vary over the surface.  The final expressions for the action of the operators can still result in rather large symbolic expressions but ultimately these can be evaluated accurately.  We use the symbolic computational package~\cite{Sympy2017} to evaluate the LHS of equation~\ref{equ_Phi_fluid} throughout our calculations.  We obtain with high precision the data needed for the method of manufactured solutions by evaluating $\mathcal{L} \bar{\Phi}$ to obtain
\begin{eqnarray}
\label{equ_RHS_Data}
\mbox{RHS Data} = \star \mb{d} \mb{b}^{\flat} = \mathcal{L} \bar{\Phi}.
\end{eqnarray}
The $\mathcal{L}$ denotes the differential operator that appears on the LHS of equation~\ref{equ_Phi_fluid}.  We mention that this would correspond to the surface force density  
$\mb{b}^{\flat} = (\bs{\delta}\mb{d})^{-1} (-\star\mb{d}) \mathcal{L} \bar{\Phi}$ in equation~\ref{equ_Stokes_geometric}.

We consider hydrodynamic flows with velocity fields generated by the vector potential
\begin{eqnarray}
\label{equ_hydro_forcing_ez}
\Phi(\mb{x}) = \frac{\exp(z)}{(4 - x)(4 - y)}.
\end{eqnarray}
Note that we use the ambient space coordinates $\mb{x} = (x,y,z)$ to avoid issues with surface coordinates that for spherical topologies would require multiple coordinate charts to describe the function on the entire surface.  

\begin{figure}[H]
\centering
\includegraphics[width=1.0\linewidth]{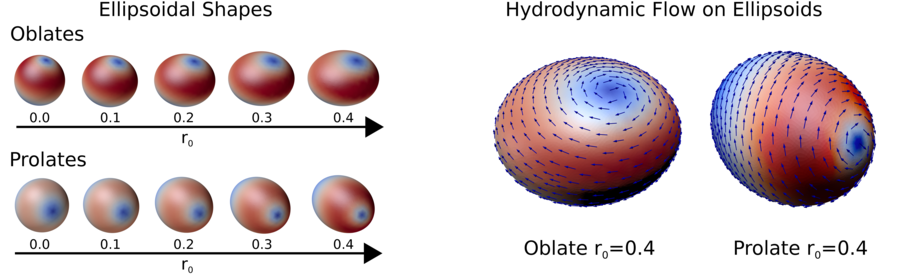} 
\caption{Hydrodynamic Flow on Ellipsoids.  We show on the left the ellipsoid shapes in the oblate and prolate cases from equation~\ref{equ_ellipsoid_shape_oblate} and~\ref{equ_ellipsoid_shape_oblate} as $r_0$ is varied.  We use these shapes for computing hydrodynamic flows driven by the surface force density in equation~\ref{equ_RHS_Data} corresponding to the vector potential of equation~\ref{equ_hydro_forcing_ez}. We investigate the accuracy of the hydrodynamics solver as the number of spherical harmonics increases and the shape is varied.  We show on the right the hydrodynamics flows corresponding to equation~\ref{equ_hydro_forcing_ez} in the case of the oblate and prolate when $r_0 = 0.4$.
}
\label{fig:ellipsoid_hydro}
\end{figure}

We solve the hydrodynamic equations~\ref{equ_Phi_fluid} using our numerical solver discussed in Section~\ref{sec_num_methods}.  We investigate how the hydrodynamics solver performs as we refine the approximation by increasing the number of spherical harmonics.  We also investigate how the convergence behaves when we vary the shape of the manifold.  We do this for ellipsoids having the prolate and oblate shapes given by equation~\ref{equ_ellipsoid_shape_oblate} and equation~\ref{equ_ellipsoid_shape_prolate} as
shown in Figure~\ref{fig:ellipsoid_hydro}.

\begin{figure}[H]
\centering
\includegraphics[width=\linewidth]{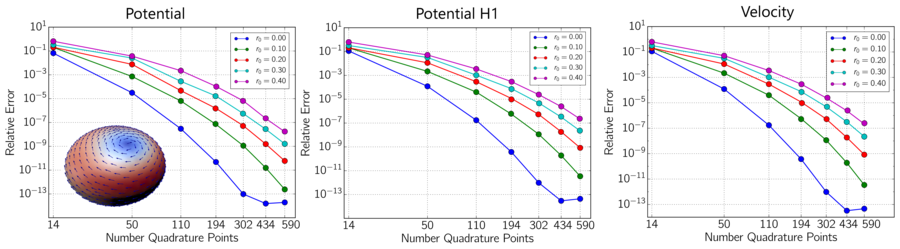} 
\caption{Convergence of the Stokes Flow for Oblate Ellipsoids.  We show the relative errors of the $L^2$-norm of the potential, $H^1$-norm of the potential, and $L^2$-norm of the velocity.  The results show how the error behaves as we increase the number of quadrature nodes $Q$ and number of spherical harmonics.  We use spherical harmonics up to degree $\lfloor L/2 \rfloor$ where $L$ is the largest exact order of the corresponding Lebedev quadrature with $Q$ nodes.  We also show how convergence depends on the shape as $r_0$ is varied.  We find in each case super-algebraic convergence of the hydrodynamic solver.  
}
\label{fig:ellipsoid_oblate_conv}
\end{figure}

\begin{figure}[H]
\centering
\includegraphics[width=\linewidth]{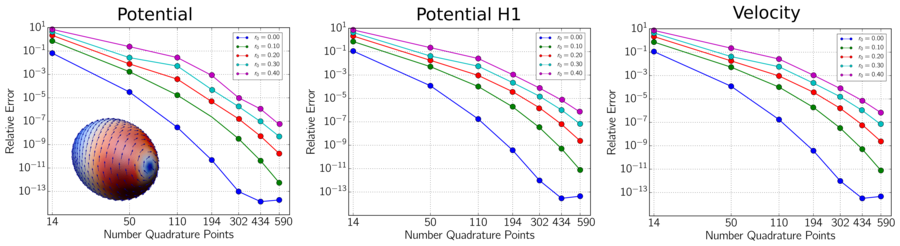} 
\caption{Convergence of the Stokes Flow for Prolate Ellipsoids.  We show the relative errors of the $L^2$-norm of the potential, $H^1$-norm of the potential, and $L^2$-norm of the velocity.  The results show how the error behaves as we increase the number of quadrature nodes $Q$ and number of spherical harmonics.  We use spherical harmonics up to degree $\lfloor L/2 \rfloor$ where $L$ is the largest exact order of the corresponding Lebedev quadrature with $Q$ nodes.  We also show how convergence depends on the shape as $r_0$ is varied.  We find in each case super-algebraic convergence of the hydrodynamic solver.  
}
\label{fig:ellipsoid_prolate_conv}
\end{figure}

We investigate as $r_0$ is varied the relative errors $\epsilon_r = \|\tilde{u} - u \|/\|u^*\|$ where $u$ is the reference solution and $\tilde{u}$ the numerical solution.  We consider both the $L^2$-norm and $H^1$-norm of the vector potential $\Phi$, and the $L^2$-norm of the velocity field $\mb{v}$.  We use on the manifold surface the $L^2$-norm given by $\|\bs{\alpha}\|^2 = \left \langle\bs{\alpha},\bs{\alpha} \right \rangle_{\mathcal{M}}$ where $\bs{\alpha}$ is a $k$-form and the $H^1$-norm is given by $\|\Phi\|_{H^1} = \left \langle\Phi,\Phi \right \rangle_{\mathcal{M}} + \left \langle \mb{d}\Phi, \mb{d} \Phi\right \rangle_{\mathcal{M}}$.  The convergence results for the numerical solver for hydrodynamics of ellipsoids when increasing the resolution of spherical harmonics and when the geometry is varied are given
for the oblate case in Figure~\ref{fig:ellipsoid_oblate_conv} and the prolate case in Figure~\ref{fig:ellipsoid_prolate_conv}.

From the convergence results in Figure~\ref{fig:ellipsoid_oblate_conv} and~\ref{fig:ellipsoid_prolate_conv}, we find in all cases that the hydrodynamics solver exhibits super-algebraic rates of convergence. We see the numerical solver can handle shapes that deviate significantly from the sphere.  For these shapes the differential operators of the hydrodynamic equations involve more complicated contributions from the geometry as seen in Appendix~\ref{appendix:radialManifold} and~\ref{appendix:diff_op_r_manifold}.  As the shapes become more pronounced we find somewhat slower convergence relative to the sphere case.  We find that for the sphere case we can capture the solution almost up to round-off error after which the errors no longer decrease.  This occurs around 434 quadrature points which exactly integrates spherical harmonics up to degree $35$ in the absence of round-off errors.

We mention that the convergence results of the velocity is especially indicative of our methods successfully capturing accurately geometric contributions.  To obtain the solution $\Phi$ to the hydrodynamics equations~\ref{equ_Phi_fluid}, this requires computing accurately the geometric contributions in the differential operators.  This includes sources of errors contributing from the Laplace-Beltrami operator, terms involving the Gaussian curvature, and also quadrature on the surface using equation~\ref{equ_surf_quad}.  To compute the fluid velocity field $\mb{v}$, this also requires computing operators with geometric contributions such as the generalized curl involving a combination of the Hodge star and exterior derivative to recover from $\Phi$ the velocity $\mb{v}$.

We notice in the results that the $H^1$ relative errors and the velocity $L^2$ relative errors are seen to be in close agreement.  This agrees with what one would intuitively expect given the close relationship between $\mb{d}\Phi$ and $\mb{v}$.  This provides another test of the accuracy of our numerical Hodge star $\star$ operator since $\mb{v}^{\flat} = -\star \mb{d}\Phi$.  This agrees with our prior work where we studied in detail a collection of numerical approximations of exterior calculus operators which we employ here on the radial manifolds~\cite{AtzbergerGross2017}.  

Many approximations enter into our solver with sources of numerical errors including the finite spherical harmonics expansions used to represent fields and the surface geometry, the surface quadratures for integration and inner-products, and the Galkerin approximation for the differential operators discussed in Section~\ref{sec_num_methods}, Appendix~\ref{appendix:radialManifold} and~\ref{appendix:diff_op_r_manifold}.  In summary, our numerical results indicate our solver provides a convergent approximation with a super-algebraic order of accuracy for surface hydrodynamics on smooth radial manifolds.

\section{Hydrodynamic Flows on Curved Fluid Interfaces}
\label{sec_results_hydro_flows}
As a demonstration of our methods, we show how our approach can be used to investigate the dependence of hydrodynamic flow responses on the geometry.  We compute flow responses motivated by particles immersed within a fluid interface and how they would move and interact through the interfacial hydrodynamic coupling.  This arises in many physical settings such as the motions of proteins within lipid bilayer membranes~\cite{HonerkampSmith2013,Powers2002,AtzbergerBassereau2014} and recent interface-embedded colloidal systems~\cite{StebePNAS2011,Choi2011,Ershov2013,DonevInterface2018,Bresme2007}.  We capture the fluid-structure coupling building on our recently introduced extended immersed boundary methods for manifolds in~\cite{AtzbergerSoftMatter2016,Peskin2002}.  While we focus here primarily on hydrodynamic flow responses, we mention that our solver could also be used as the basis for drift-diffusion simulations of microstructures on radial manifolds using our fluctuating hydrodynamics approaches~\cite{AtzbergerSELM2011,AtzbergerSIB2007,
AtzbergerLAMMPS2016,AtzbergerSoftMatter2016}.  

\begin{figure}[H]
\centering
\includegraphics[width=0.9\linewidth]{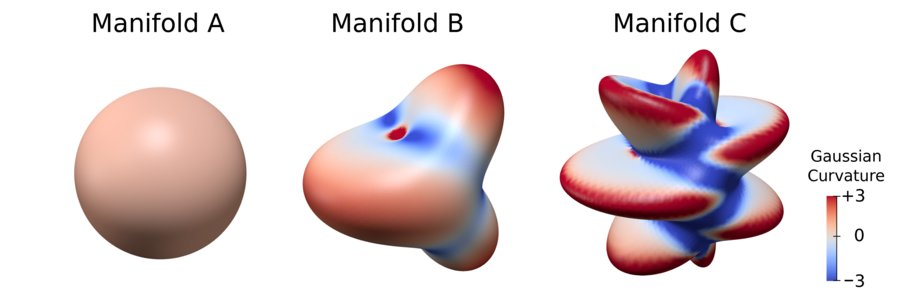}
\caption{Radial Manifold Shapes.  We consider hydrodynamic flows on manifolds with shapes ranging from the sphere to the more complicated geometries generated by equation~\ref{equ_manifold_B}.  We show with colors the Gaussian curvature of the shapes.  We take Manifold A to be a sphere of radius $R = 1.0$.  We show Manifold B with $r_0 = 0.4$ and Manifold C with $r_0 = 0.4$ in equation~\ref{equ_manifold_B}.
}
\label{fig:manifold_shapes}
\end{figure}

\begin{figure}[H]
\centering
\includegraphics[width=0.99\linewidth]{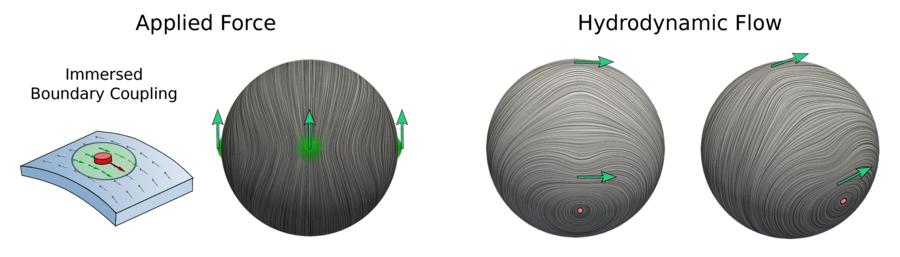}
\caption{Structure of the hydrodynamic flow.  We consider the case of three particles immersed within an interface of spherical shape and subjected to force.  We generate a force density on the surface using the extended immersed boundary method for manifolds we introduced in~\cite{AtzbergerSoftMatter2016}.  The particles are configured at the locations $\mb{x}_1 = (-1,0,0)$, $\mb{x}_2 = (1,0,0)$, $\mb{x}_3 = (0,-1,0)$, and each subjected to the force $\mb{F} = (0,0,1)$.  We show in the left panel the immersed boundary approach for fluid-particle coupling on manifolds~\cite{AtzbergerSoftMatter2016} and range of spreading around each particle used to obtain a force density on the surface.  We show in the right panel the hydrodynamic flow response.  The flow exhibits two two vortices and global recirculation of the fluid.  We visualize the streamlines of the hydrodynamics flows using Line Integral Convolution (LIC)~\cite{Cabral1993}.  Hydrodynamic results are for the case with $\mu = 0.1$,$\gamma = 0.1$ and $Q = 5810$. }
\label{fig:Stokes_LIC_Sphere}
\end{figure}

We consider the case of three particles subject to force when immersed within curved fluid interfaces having the shapes in Figure~\ref{fig:manifold_shapes}.  We generate a force density on the surface using our extended immersed boundary methods for manifolds introduced in~\cite{AtzbergerSoftMatter2016}.  In the reference spherical shape, the particles are configured at the locations $\mb{x}_1 = (-1,0,0)$, $\mb{x}_2 = (1,0,0)$, $\mb{x}_3 = (0,-1,0)$ with each subjected to the force $\mb{F} = (0,0,1)$.  For each radial manifold shape we use the push-forward of the three locations $\mb{x}_1, \mb{x}_2, \mb{x}_3$ and apply force using $\mb{F}$ projected in the tangential direction of the surface and normalized to maintain a unit force magnitude for all shapes.  We spread forces over the length-scale $0.1$ on the surface using our extended immersed boundary method discussed in~\cite{AtzbergerSoftMatter2016}.  We show the particle configuration, force density, and hydrodynamic flow response on the sphere in Figure~\ref{fig:Stokes_LIC_Sphere}.

We remark that throughout our numerical calculations we allow for a net total force $F_T$ or torque $\tau_T$ acting on the manifold which physically could drive rotational and translational rigid body motions of the entire interface within the surrounding fluid.  We resolve explicitly the tangential contributions of the rigid-body motions with our numerical solver, and treat implicitly the non-tangential contributions.  Of course the flow field in other reference frames can be recovered by adding the non-tangent $\mb{v}_0$ velocity field at each location on the surface.  

We consider the radial manifold shapes shown in Figure~\ref{fig:manifold_shapes}.  These shapes are generated by the radial functions $r(\theta,\phi)$.  For Manifold A which is a sphere we have $r(\theta,\phi) = 1.0$.  For Manifold B and C we use 
\begin{eqnarray}
\label{equ_manifold_B}
\label{equ_manifold_C}
r(\theta,\phi) = 1 + r_0\sin(3\phi)\cos(\theta) \mbox{\hspace{0.2cm}(Manifold B)}, \hspace{0.5cm}
r(\theta,\phi) = 1 + r_0 \sin(7\phi)\cos(\theta) \mbox{\hspace{0.2cm}(Manifold C)}.
\end{eqnarray}
Additional details concerning the differential geometry of these radial manifolds can be found in our prior work~\cite{AtzbergerGross2017} and in Appendix~\ref{appendix:radialManifold} and~\ref{appendix:diff_op_r_manifold}.

\begin{figure}[H]
\centering
\includegraphics[width=1.0\linewidth]{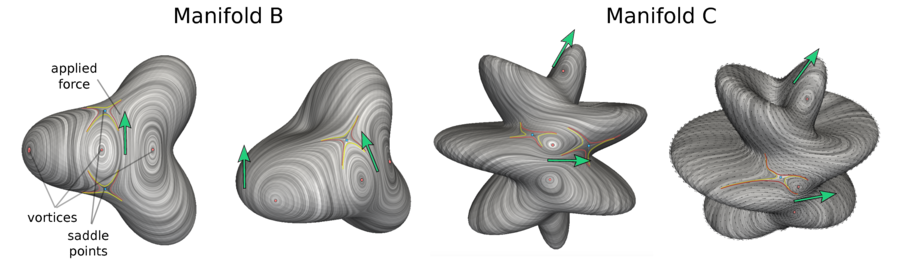}
\caption{Structure of the Hydrodynamic Flow.  We show for Manifold B and C the hydrodynamic flow responses for a localized unit force applied in the tangential direction to particles on the surface at the push-forward of the locations discussed in Figure~\ref{fig:Stokes_LIC_Sphere}.  We visualize the streamlines using Line Integral Convolution (LIC)~\cite{Cabral1993}.  The flow responses exhibit eight critical points corresponding to six vortices and four saddle points.  The vortices are marked with red points and the saddle points with cyan points.  We sketch approximate separatrices for each of the saddle points.  For these surfaces the hydrodynamic flows appear to exhibit structures that favor more localized recirculation of the fluid relative to the responses seen for the sphere in Figure~\ref{fig:Stokes_LIC_Sphere}.  }
\label{fig:Stokes_LIC_dimpleAndFountain1}
\end{figure}

We demonstrate how the solver captures geometric effects when varying shapes transitioning from a sphere to either Manifold B or Manifold C. {Since our manifolds are always compact with spherical topology, we have from the Poincare-Hopf Theorem that our surface flows must have singularities~\cite{JostBookDiffGeo1991,SpivakDiffGeo1999}.}  We consider hydrodynamic flow responses on shapes when changing the amplitude $r_0$ in the range $0.0$ to $0.4$ in equation~\ref{equ_manifold_B}.  We find for shapes having a relatively homogeneous curvature close to a sphere that the flows are observed to recirculate the interfacial fluid globally with just two vortices as in Figure~\ref{fig:Stokes_LIC_Sphere}.  Interestingly, as the shapes become more complicated with heterogeneous positive and negative curvatures we see that the flow responses are observed to undergo quantitative changes and a topological transition exhibiting the creation of new vortices and saddle-point stagnation points as seen in Figure~\ref{fig:Stokes_LIC_dimpleAndFountain1}.  {This appears to arise from the hydrodynamic flow recirculating fluid more locally and from rigid-body rotational motions of the interface, which we discuss more below and in Appendix~\ref{appendix:rigidBodyRotation}}.  To investigate this further, we characterize the hydrodynamics and contributions of geometry by quantifying the dissipation rates associated with flows on curved surfaces.

Stokes hydrodynamics can be viewed as solving a variational principle through the Helmholtz Minimum Dissipation Theorem~\cite{batchelor2000,Helmholtz1868}.  This corresponds to the flow minimizing the Rayleigh-Dissipation in the space of solenoidal velocity fields subject to boundary or auxiliary conditions~\cite{Gelfand2000}.  We generalize this result and the Rayleigh-Dissipation rate to obtain a variational principle $\mbox{Q}[\mb{v}^{\flat}]$ for the Stokes hydrodynamics equation~\ref{equ_Stokes_geometric} for curved fluid surfaces.  This can be expressed using the exterior calculus as 
\begin{eqnarray}
\label{equ:var_prob}
&& \inf_{\mb{v}} \mbox{Q}[\mb{v}^{\flat}], \mbox{\hspace{0.0cm} where} \\
\label{equ:Q_rate}
\mbox{Q}[\mb{v}^{\flat}] & = & \mbox{RD}[\mb{v}^{\flat}] - \mbox{F}[\mb{v}^{\flat}] \\ 
\label{equ:RD_rate}
\mbox{RD}[\mb{v}^{\flat}] & = & \mu_m \langle \mb{d}\mb{v}^{\flat}, \mb{d}\mb{v}^{\flat} \rangle_\mathcal{M} - 2\mu_m \langle K\mb{v}^{\flat},\mb{v}^{\flat}  \rangle_\mathcal{M} + \gamma \langle \mb{v}^{\flat},\mb{v}^{\flat} \rangle_\mathcal{M} \\
\label{equ:F_rate}
\mbox{F}[\mb{v}^{\flat}] & = & \langle \mb{b}^{\flat},\mb{v}^{\flat} \rangle_\mathcal{M}.
\end{eqnarray}
Here the minimization in $\mb{v}^{\flat}$ is constrained to be over smooth solenoidal vector fields on the surface in the sense $-\bs{\delta} \mb{v}^{\flat} = 0$ and $\mb{v}^{\flat} \in H^{2}(\mathcal{M})$~\cite{Lieb2001,Acheson1990}.  For the given fluid constitutive laws, the Rayleigh-Dissipation term $\mbox{RD}[\mb{v}^{\flat}]$ of equation~\ref{equ:RD_rate} gives the rate at which the fluid is doing work when flowing according to the velocity field $\mb{v}^{\flat}$.  The term $\mbox{F}$ of equation~\ref{equ:F_rate} corresponds to the work done by the external body forces acting to drive the fluid.  We derived equation~\ref{equ:Q_rate} by taking the manifold inner product of $\mb{v}^{\flat}$ with both sides using equation~\ref{equ_Stokes_geometric}. Central to our derivation is use of the adjoint property of the exterior derivative $\mb{d}$ with the co-differential $\bs{\delta}$ in the sense of equation~\ref{equ_adjoint_d_delta}.  It can be shown that the solution of the Stokes equations~\ref{equ_Stokes_geometric} minimizes equation~\ref{equ:Q_rate} over all velocity fields subject to the incompressibility constraint $-\bs{\delta} \mb{v}^{\flat} = 0$.  Taking variational derivatives of equation~\ref{equ:Q_rate} it readily follows that the Stokes equations~\ref{equ_Stokes_geometric} are recovered as the Euler-Lagrange equations of the variational problem given in equation~\ref{equ:var_prob}~\cite{Gelfand2000}.  The constraints when handled by the method of Lagrange multipliers~\cite{Gelfand2000} gives the pressure term in equations~\ref{equ_Stokes_geometric}.  

The variational principle in equation~\ref{equ:Q_rate} provides a useful way to view the hydrodynamic flows as arising from competing physical effects.  For steady flows there is a balance between the work done by an external body force with the dissipation from the solvent drag and the dissipation from the internal shearing motions of the fluid.  We see the dissipation from shearing motions can be split into two parts. The first term in equation~\ref{equ:RD_rate} is equivalent to
$\mu_m \langle \mb{d}\mb{v}^{\flat}, \mb{d}\mb{v}^{\flat} \rangle_\mathcal{M} = \mu_m \langle -\star\mb{d}\mb{v}^{\flat}, -\star\mb{d}\mb{v}^{\flat} \rangle_\mathcal{M} = \mu_m \langle \mbox{curl}_{\mathcal{M}}\mb{v}^{\flat}, \mbox{curl}_{\mathcal{M}}\mb{v}^{\flat} \rangle_\mathcal{M} = \mu_m \| \mbox{curl}_{\mathcal{M}}\mb{v}^{\flat} \|_{\mathcal{M}}^2$. This corresponds to creation of vorticity $\omega = \mbox{curl}_{\mathcal{M}}\mb{v}^{\flat}$ within the fluid.  

The second term involves the Gaussian curvature $K$ which depending on the sign either penalizes or promotes relative to the flat case changes in the magnitude of the fluid velocity.  Since these two terms arise from the shearing motions of the fluid material in the ambient space they have a strong relationship through the surface geometry.  For regions having the same vorticity distribution $\omega$ the curvature weighted term shows that regions with positive Gaussian curvature have a smaller rate of dissipation relative to regions having negative Gaussian curvature.  We see that unlike the flat case the local curvature of the surface requires the fluid to flow with a momentum in the ambient space that must change locally in direction to remain within the surface.  As a consequence, we see for surface constrained flows the geometry can result in additional sources of shearing motions and dissipation.  

The third term corresponds to dissipation from drag of the surface fluid with the surrounding solvent fluid depending in this case only on the total manifold $L^2$-norm of the flow.  The term $\mbox{F}[\mb{v}^{\flat}]$ associated with equation~\ref{equ:F_rate} corresponds to the applied body force and penalizes the flow when it is not aligned with the force density $\mb{b}$.  We see that expressing the Rayleigh-Dissipation with exterior calculus and splitting into these distinct terms starts to reveal some of the interesting ways hydrodynamic responses can depend on the geometry of the surface. 

\begin{figure}[H]
\centering
\includegraphics[width=0.48\linewidth]{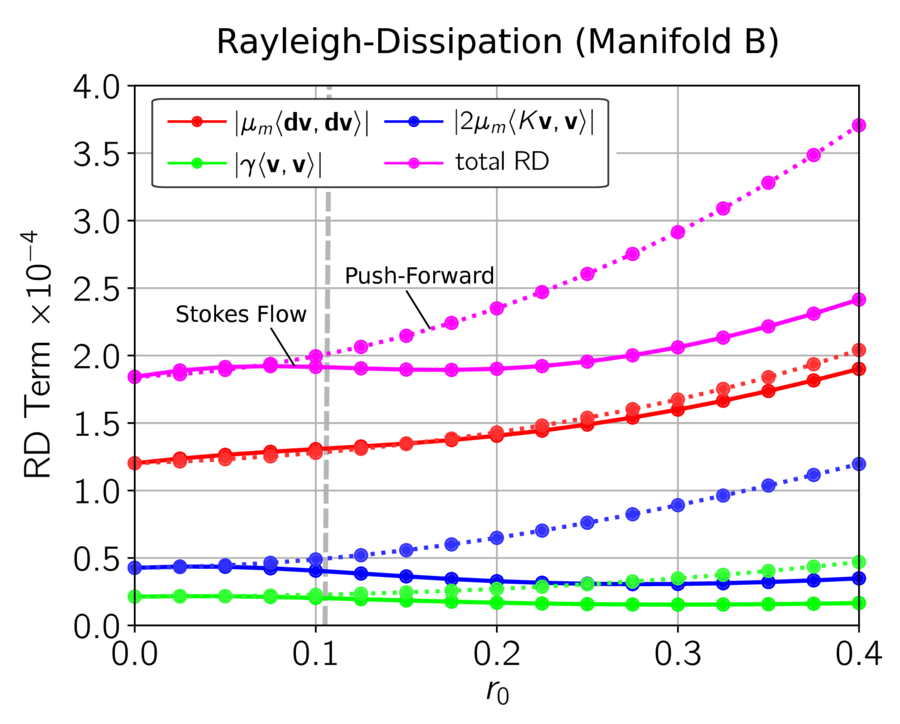}
\includegraphics[width=0.48\linewidth]{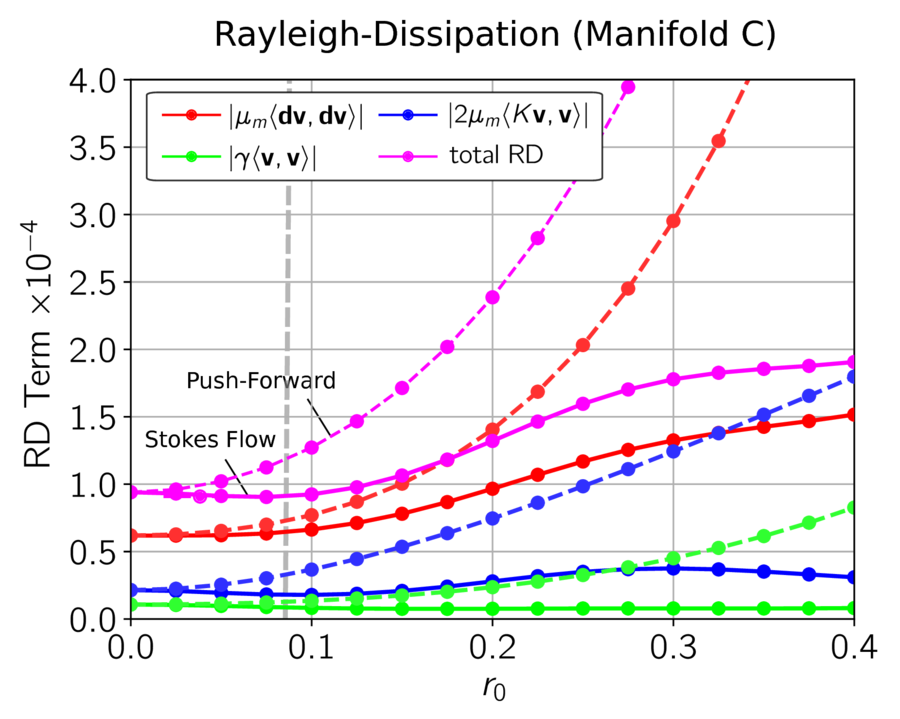}
\caption{Rayleigh-Dissipation Rates of Hydrodynamic Flows.  Hydrodynamic flows $\mb{v}$ on surfaces are obtained by solving the Stokes equations~\ref{equ_Stokes_geometric} as $r_0$ is varied.  We take $\mu_m = 0.1$ and $\gamma = 0.1$ for the manifolds in Figure~\ref{fig:manifold_shapes}.  RD rates for $\mb{v}$ are shown as solid curves.  For comparison we consider the rates obtained from the velocity field $\hat{\mb{v}}$ of flow on the sphere (case $r_0 = 0$) obtained by the pushed-forward $\mb{\tilde{v}} = \phi_* \hat{\mb{v}}$ to the manifold shape with given $r_0$.  RD rates for $\mb{\tilde{v}}$ are shown as dotted curves.  We find that as the geometry deviates from the sphere the rates for the Stokes flow on the manifold become significantly smaller than the push-forward flow fields from the sphere.  We find the two cases begin to diverge significantly in the regime where the velocity field transitions topologically with the addition of new vortices and saddle-point stagnation points as in Figure~\ref{fig:Stokes_LIC_dimpleAndFountain1}.  This transition occurs for Manifold B around $r_0^* = 0.105$ and for Manifold C around $r_0^* = 0.085$ (vertical dashed line).  These results indicate some of the ways that surface geometry can contribute to dissipation rates and hydrodynamic flow responses.
}
\label{fig:RayleighDissipation}
\end{figure}

We compute the dissipation rates for hydrodynamic flow responses for the geometries ranging from the sphere to Manifold B and C when varying $r_0$ in the range $0.0$ to $0.4$ in equation~\ref{equ_manifold_B}.  We show the final shapes in Figure~\ref{fig:shapes}.  As a basis for comparison for investigating the role of the geometry we consider both the Stokes hydrodynamic response $\mb{v}$ obtained from our solver for equation~\ref{equ_Stokes_geometric} and the flow obtained from the push-forward of the flow $\hat{\mb{v}}$ from the sphere to the surface geometry $\tilde{\mb{v}} = \phi_* \hat{\mb{v}}$.  The $\phi$ denotes the radial mapping from the sphere to the surface geometry and $\phi_*$ the associated push-forward~\cite{Abraham1988}.  We show our results for the dissipation rates for hydrodynamic flow responses $\mb{v}$ and push-forward flows $\tilde{\mb{v}}$ in Figure~\ref{fig:RayleighDissipation}.

\begin{figure}[H]
\centering
\includegraphics[width=0.95\linewidth]{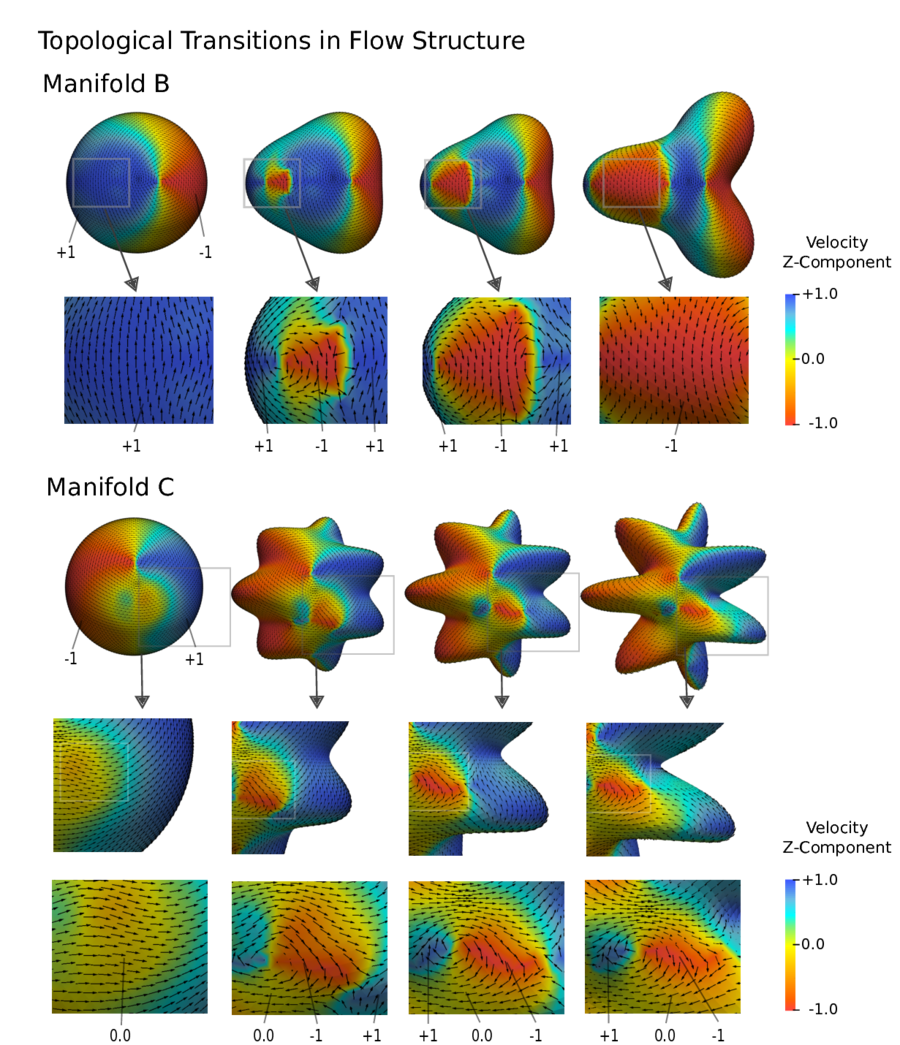}
\caption{Topological Transitions in the Flow Structure.  As the shapes of the manifolds deviate more from the sphere the velocity field of the hydrodynamic flow undergoes a topological transition with the creation of new vortices and saddle-points.  The topological structures appear to correspond with the flow reorganizing to recirculate fluid in a more localized manner relative to the global recirculation seen on the sphere.  This is especially pronounced in the elongated geometries that form for the Manifold C shapes.  We show configurations for Manifold B and Manifold C when $r_0 = 0.0, 0.15, 0.25, 0.4$ in equation~\ref{equ_manifold_B}.  We quantify the Rayleigh-Dissipation associated with each of these flows in Figure~\ref{fig:RayleighDissipation}.}
\label{fig:Stokes_flows_geoRole_dimple}
\label{fig:Stokes_flows_geoRole_fountain}
\end{figure}

We find as we transition from the sphere to Manifold B and C with the shapes becoming more heterogeneously curved the dissipation rates for the Stokes flow is significantly smaller than the push-forward flow $\tilde{\mb{v}}$.  The differences become especially large after the Stokes flows exhibit the topological transition with the emergence of new vortices and saddle-points as seen in Figure~\ref{fig:Stokes_LIC_dimpleAndFountain1}.  We see as the geometry is varied in the range of $r_0$ around the topological transitions the Rayleigh-Dissipation appears to remain relatively constant in both cases in Figure~\ref{fig:RayleighDissipation}.  This seems to indicate that changes in the flow can accommodate to some extent changes in the geometry to avoid the otherwise increases in dissipation that would have occurred within the fluid interface if remaining with the flow structure associated with the sphere case.  We show the flow responses and relation to geometry in more detail in the plots of Figure~\ref{fig:Stokes_flows_geoRole_dimple}.  The geometry appears to promote for both Manifold B and C a recirculation of the fluid more locally to regions of positive Gaussian curvature possibly at the expense of creating some additional local vorticity in the fluid flow, see Figure~\ref{fig:Stokes_LIC_dimpleAndFountain1} and Figure~\ref{fig:Stokes_flows_geoRole_dimple}.  {We also emphasize that the quantitative and topological changes can also in part be explained by the generation from the forces acting on the fluid interface body that result in a rigid-body rotational motion.  We show results with the rigid-body rotational motions subtracted in Appendix~\ref{appendix:rigidBodyRotation}.  We see that in the case of the sphere we can obtain similar locally re-circulating flows when viewed in the moving reference-frame of the rotating fluid interface.}

In our numerical results for Manifold B and C, we see that our geometries can exhibit elongated regions of positive Gaussian curvature surrounded by regions of negative Gaussian curvature.  This is especially prominent for Manifold C as seen in Figure~\ref{fig:shapes} and~\ref{fig:Stokes_flows_geoRole_dimple}.  The role of the negative Gaussian curvature term in equation~\ref{equ:RD_rate} indicates that dissipation rates can increase relative to regions of positive curvature.  This could potentially explain the preference of the fluid to recirculate more locally to avoid having to flow through regions of negative curvature.  Related to our findings, there has also been some related work concerning fluid dissipation in the case of perturbations from a flat sheet, as arise in planar soap films, which were recently reported in~\cite{Debus2017}.  These authors also find that curvature-induced dissipation can amplify dissipation and affect structure of the hydrodynamic flow.  In our work we see even further phenomena with observed topological transitions in the structure of the hydrodynamic flow response.  

{We show that changes in the observed hydrodynamic responses can in part be explained by the rigid-body rotational motions induced by the non-torque balance force acting on the fluid interface seen in the ambient space reference frame.  We show the hydrodynamic velocity field in the moving reference frame counter-rotated by the rigid-body rotational motions in Appendix~\ref{appendix:rigidBodyRotation}.  We see in the case of the sphere we have similar vortices and saddle points when observed in the moving reference frame.  In either reference frame, we find the hydrodynamic flow responses favor quantitatively more localized re-circulation of fluid.  These results indicate some of the rich mechanics and related phenomena captured by our solver that can arise when going beyond the often considered setting of infinite fluid sheets to instead consider hydrodynamic flows confined within compact geometries.}

The numerical results we have present here to demonstrate our hydrodynamics solver indicate some of the rich ways geometry could have implications for hydrodynamic coupling and possible kinetic consequences for the motions of inclusion particles immersed in curved fluid interfaces.  Further investigations also could be performed readily into the role of geometry in surface hydrodynamic phenomena using our introduced solver.  {We also mention that our solver can be used as the basis for developing extended immersed boundary methods for manifolds~\cite{AtzbergerSoftMatter2016} and drift-diffusion simulations of particles and microstructures within curved fluid interfaces building on our fluctuating hydrodynamics approaches~\cite{AtzbergerSELM2011,AtzbergerSIB2007,AtzbergerSoftMatter2016}.  These approaches could be useful in computing interfacial mobilities and surface kinetics for many systems, such as proteins within curved lipid bilayer membranes, polymeric networks in cell biology like the spectrin network of the red-blood cell, or self-assembly for colloidal systems immersed in fluid interfaces.  We also expect many of the underlying ideas used in our solver could be used to develop solvers for other PDEs on radial manifolds.}

\section{Conclusion}
\label{sec_conclusions}
We have developed spectrally accurate numerical methods for solving hydrodynamic flows on radial manifolds.  We have formulated hydrodynamic equations using approaches from continuum mechanics in the manifold setting and exterior calculus.  We developed spectral numerical methods for surface hydrodynamics on radial manifold shapes using  Galerkin approximation and hyper-interpolation of spherical harmonics.  We find that surface geometry can play a significant role in hydrodynamic flow responses.  We gained some insights into geometric contributions by considering the Rayleigh-Dissipation rates of the Stokes flows as the surface geometry was varied.  We found that for manifolds having a geometry that exhibit heterogeneous positive and negative curvatures of sufficient amplitude the hydrodynamic responses can exhibit interesting quantitative changes and topological transitions.  These transitions manifest with the creation of new flow structures such as the creation of vortices or saddle-points.  {We showed these observations can in part be explained by the rigid-body rotational motions of the fluid interface when viewed from the reference frame of the ambient space.}  These results indicate some of the rich physical phenomena that can occur within hydrodynamic flows on curved surfaces captured by our methods.  We mention that our numerical solvers could be useful as the basis for the further development of extended immersed boundary methods for curved fluid interfaces~\cite{AtzbergerSoftMatter2016,AtzbergerPadidar2018}.  {Also our methods can be used for investigating the drift-diffusion dynamics of particles and microstructures using our fluctuating hydrodynamics approaches~\cite{AtzbergerSoftMatter2016,AtzbergerSELM2011,AtzbergerSIB2007}.  We expect these approaches to be useful in studies of diverse physical systems where curvature plays a role.  Potential applications include studies of the kinetics of curvature inducing proteins within lipid bilayer membranes, membrane associated polymers like the spectrin network in red blood cells, or the self-assembly of colloids within curved interfaces.}  Many of the ideas underlying our numerical methods can also be adopted for solving other partial differential equations on radial manifolds.  

\section{Acknowledgments}
\label{sec_acknowledgements}
We would like to acknowledge support to P.J.A. and B.J.G. from research grants DOE ASCR CM4 DE-SC0009254, NSF CAREER Grant DMS-0956210, and NSF Grant DMS - 1616353.  We would also like to thank the Kozatos for generous support for B.J.G.  We also acknowledge the UCSB Center for Scientific Computing NSF MRSEC (DMR-1121053) and UCSB MRL NSF CNS-0960316.  

\bibliographystyle{plain}
\bibliography{paperDatabase}{}

\appendix

\section*{Appendix}

\section{Role of Rigid-Body Rotational Motions}
\label{appendix:rigidBodyRotation}

{We investigate the hydrodynamic responses to a non-torque balanced force density applied to the fluid interface.  For our compact manifolds,  this can result in a rigid-body rotational motion within the ambient fluid.  We investigate here the role of the rigid-body rotational motions on the observed hydrodynamic responses and velocity fields.  When a force is applied the fluid interface responds with a combination of localized internal flows and global rigid-body rotation.}

{The trade-off between these global and local responses is governed by the traction stress with the surrounding bulk fluid.  In equation~\ref{equ_v_stokes}, we used a basic drag model with traction stress $\bs{\tau}_f = -\gamma \mb{v}$.  For a fully hydrodynamic approach, which would require a solver for the bulk fluid flow, the traction stress would be $\bs{\tau}_f = \mu_{f} \left(\nabla \mb{v} + \nabla \mb{v}^T\right)\cdot\mb{n}$.  We can relate these parameters for the purpose of obtaining scaling relations by $\mu_f \sim \gamma \ell_f$, where $\ell_f$ is a characteristic length-scale associated with variations in the flow field.  We can characterize the expected relative strength of the external traction stress with the internal shear stresses of the interface by the Saffman-Delbr\"{u}ck (SD) length ratio $L/R$~\cite{AtzbergerSoftMatter2016,Saffman1975}.  The SD ratio characterizes the length-scale over which the interfacial flow field varies in response to a point force, where $L = \mu_m/2\mu_f \sim \mu_m/2\gamma\ell_f$ and $R$ is the effective radius of the manifold~\cite{AtzbergerSoftMatter2016}.  These considerations suggest that when $L/R \ll 1$ the external traction stress $\bs{\tau}_f$ becomes large relative to the internal shear stresses and the hydrodynamic responses avoid rigid rotations and prefer instead to have flows that are more localized within the fluid interface.  When $L/R \gg 1$, the external traction stress $\bs{\tau}_f$ becomes small relative to internal shear stresses and the hydrodynamic responses prefer now to rotate the entire interface rigidly within the bulk fluid with relatively less internal hydrodynamic flows that would result in intra-interfacial shear stresses.}

\begin{figure}[H]
\centering
\includegraphics[width=0.8\linewidth]{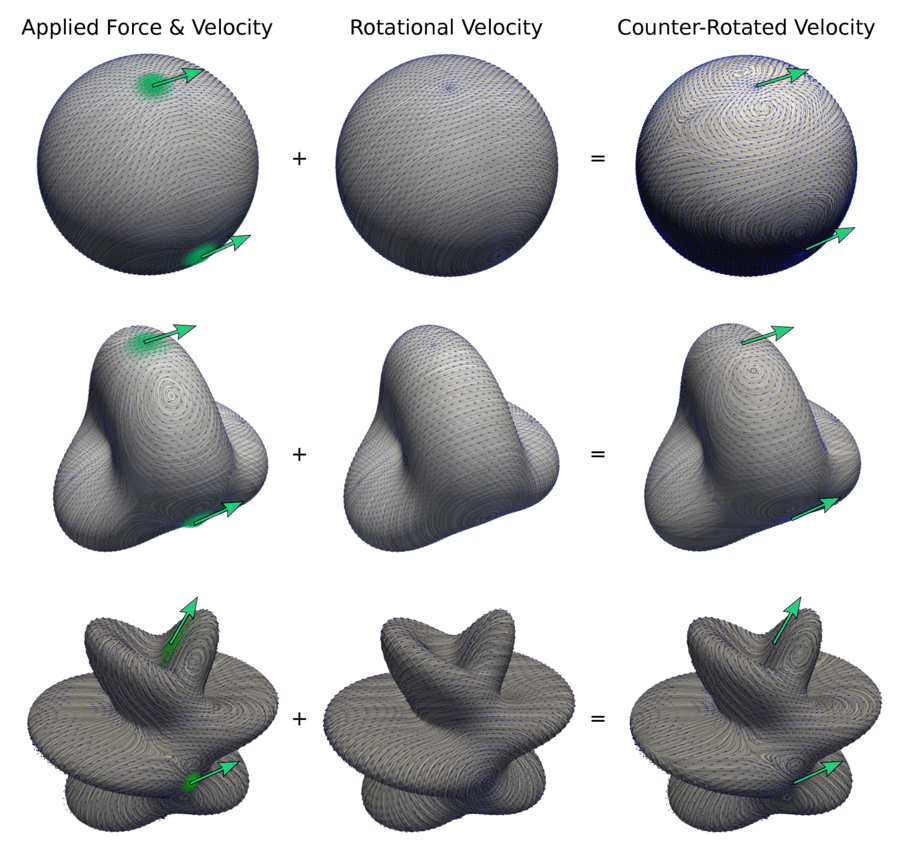}
\caption{Counter-Rotated Hydrodynamic Flows.  We consider the case of three particles configured at the locations $\mb{x}_1 = (-1,0,0)$, $\mb{x}_2 = (1,0,0)$, $\mb{x}_3 = (0,-1,0)$ embedded within an interface and subjected to force $\mb{F} = (0,0,1)$ for the sphere and tangent for the other manifolds.  We use our extended immersed boundary method for manifolds we introduced in~\cite{AtzbergerSoftMatter2016} and Section~\ref{sec_results_hydro_flows}.  In the left panel, we show the hydrodynamic responses for $\mu = 0.1$,$\gamma = 0.1$ and $Q = 5810$.  In the middle panel, we show the rotational field that best counters the rigid-body rotation of the fluid interface.  In the right panel, we have combining the velocity fields to arrive at a counter-rotated velocity field that would be observed in the moving reference frame.  We visualize the streamlines of the hydrodynamics flows using Line Integral Convolution (LIC)~\cite{Cabral1993}.}
\label{fig:Sphere_CounterRotated}
\end{figure}

We investigate hydrodynamic responses by performing a study that finds for each hydrodynamic flow the best approximating rigid-body rotational motion and then look at the counter-rotated velocity so the rotational motion can be subtracted from the hydrodynamic velocity field.  This conversion of the velocity corresponds to making observations in a moving reference frame that rotates in agreement with the rigid rotational motion of the fluid interface.  We show the flows in this moving reference frame for the spherical case (Manifold A) and Manifold B and Manifold C in Figure~\ref{fig:Sphere_CounterRotated}.

{We find the rigid-body rotational motion when subtracted from the hydrodynamic response results within the interface in a localize flow pattern for the sphere case (Manifold A), see right-panel of Figure~\ref{fig:Sphere_CounterRotated}.  We see this counter-rotation has less of a qualitative impact on the hydrodynamic responses for Manifold B and C.  This in part arises since the complex shapes have larger surface area and thus larger external traction stresses that further inhibit rigid-body rotational motions relative to internal flows.  These results have important implications for how the hydrodynamic flow fields are to be interpreted depending on the circumstances and reference-frame of interest in a given problem or application.  When considering internal mixing or transport within the hydrodynamic interface itself, or when the interface is immobilized to prevent it from rotating, the interpretation in the reference-frame associated with the rigid-body rotational motion of the fluid interface may be the most appropriate.  When considering laboratory measurements or how outside entities in the bulk fluid, such as particles or polymers, interact with the moving interface the rotational motions may play a significant role and the reference frame in the ambient space may be most appropriate.  Our numerical solvers can be utilized in either of these cases to resolve the hydrodynamic flow responses.  As we see from Figure~\ref{fig:Sphere_CounterRotated}, these rigid-body rotational motions can in part account for the quantitative and topological transitions in stream-lines exhibited from the perspective of the ambient space reference-frame.  These results show for compact geometries the importance of considering the role of the rigid-body rotational motions in flow responses captured by our solvers and in the mechanics of the interfacial hydrodynamics we developed in Section~\ref{sec_cont_mechanics}.}

\section{Differential Geometry of Radial Manifolds}
\label{appendix:radialManifold}
Radial manifolds have a shape that is defined by a surface where each point can be connected by a line segment to the origin without intersecting the surface.  We can express radial manifolds shapes in spherical coordinates ($\theta,\phi)$ as the collection of points $\mb{x}$ of the form
\begin{eqnarray}
\label{equ:manifoldParam}
\mb{x}(\theta,\phi) = \bs{\sigma}(\theta, \phi) =  r(\theta, \phi)\mb{r}(\theta,\phi).
\end{eqnarray}
The $\mb{r}$ is the unit vector from the origin to the point on the sphere corresponding to angle $\theta,\phi$ and $r$ is a positive scalar function.

We take an isogeometric approach to representing the manifold $M$.  We sample the scalar function $r$ at the Lebedev nodes and represent the geometry using the finite spherical harmonics expansion $r(\theta,\phi) = \sum_i \bar{r}_i Y_i$ up to the order $\lfloor L/2 \rfloor$ where $\bar{r}_i = \langle r, Y_i \rangle_Q$ for a quadrature of order $L$.  We discuss spherical harmonics in Appendix~\ref{appendix:sphericalHarmonics}.

We consider two coordinate charts for our calculations.  The first is referred to as Chart A and has coordinate singularities at the north and south pole.  The second is referred to as Chart B and has coordinate singularities at the east and west pole~\cite{AtzbergerSoftMatter2016}.  For each chart we use spherical coordinates with $(\theta, \phi) \in [0 , 2\pi) \times [0 , \pi]$ but to avoid singularities only use values in the restricted range $\phi \in [\phi_{min}, \phi_{max}]$, where $ 0 < \phi_{min} \le \frac{\pi}{4}$, and $\frac{3\pi}{4} \le \phi_{max} < \pi$.  In practice, one typically takes $\phi_{min} = 0.8\times \frac{\pi}{4}$ and $\phi_{max} = 0.8 \times \pi$. 
For Chart A, the manifold is parameterized in the embedding space $\mathbb{R}^3$ as
\begin{eqnarray}
\mb{x}(\hat{\theta}, \hat{\phi}) = r(\hat{\theta}, \hat{\phi})\mathbf{r}(\hat{\theta}, \hat{\phi}), \hspace{0.5cm} 
\mathbf{r}(\hat{\theta}, \hat{\phi}) = \begin{bmatrix}\sin(\hat{\phi})\cos(\hat{\theta}), & \sin(\hat{\phi})\sin(\hat{\theta}), & \cos(\hat{\phi}) \end{bmatrix} 
\label{equ:defChartA}
\end{eqnarray}
and for Chart B 
\begin{eqnarray}
\mb{x}(\bar{\theta}, \bar{\phi}) = r(\bar{\theta}, \bar{\phi})\mathbf{r}(\bar{\theta}, \bar{\phi}), \hspace{0.5cm} 
\mathbf{\bar{r}}(\bar{\theta}, \bar{\phi}) = \begin{bmatrix} \cos(\bar{\phi}), & \sin(\bar{\phi})\sin(\bar{\theta}), & -\sin(\bar{\phi})\cos(\bar{\theta}) \end{bmatrix}.
\label{equ:defChartB}
\end{eqnarray}
With these coordinate representations, we can derive explicit expressions for geometric quantities associated with the manifold such as the metric tensor and shape tensor.  The derivatives used as the basis $\partial_\theta, \partial_\phi$ for the tangent space can be expressed as
\begin{eqnarray}
\label{eqn:sigmaTheta}
\bs{\sigma}_{\theta}(\theta, \phi) & = & r_{\theta}(\theta, \phi)\mb{r}(\theta, \phi) + r(\theta, \phi) \mb{r}_{\theta}(\theta, \phi)\\
\label{eqn:sigmaPhi}
\bs{\sigma}_{\phi}(\theta, \phi) & = & r_{\phi}(\theta, \phi)\mb{r}(\theta, \phi) + r(\theta, \phi) \mb{r}_{\phi}(\theta, \phi).
\end{eqnarray}
We have expressions for $\mb{r}_{\theta}$ and $\mb{r}_{\phi}$ in the embedding space $\mathbb{R}^3$ using equation~\ref{equ:defChartA} or equation~\ref{equ:defChartB} depending on the chart being used.  The first fundamental form $\mathbf{I}$ (metric tensor) and second fundamental form $\mathbf{II}$ (curvature tensor) are given by 
\begin{align}
\mathbf{I} = \begin{bmatrix}
E & F \\
F & G
\end{bmatrix} = \begin{bmatrix}
\bs{\sigma}_{\theta} \cdot \bs{\sigma}_{\theta} & \bs{\sigma}_{\theta} \cdot \bs{\sigma}_{\phi} \\
\bs{\sigma}_{\phi} \cdot \bs{\sigma}_{\theta} & \bs{\sigma}_{\phi} \cdot \bs{\sigma}_{\phi}
\end{bmatrix} =  \begin{bmatrix}
r_{\theta}^2+r^2\sin(\phi)^2 & r_{\theta}r_{\phi} \\
r_{\theta}r_{\phi} & r_{\phi}^2+r^2
\end{bmatrix}. \label{equ:I_mat}
\end{align}
and 
\begin{align}
\mathbf{II} = \begin{bmatrix}
L & M \\
M & N
\end{bmatrix} = \begin{bmatrix}
\bs{\sigma}_{\theta \theta} \cdot \bs{n} & \bs{\sigma}_{\theta \phi} \cdot \bs{n} \\
\bs{\sigma}_{\phi \theta} \cdot \bs{n} & \bs{\sigma}_{\phi \phi}\cdot \bs{n}
\end{bmatrix}. \label{equ:II_mat} 
\end{align}
The $\mb{n}$ denotes the outward normal on the surface and is computed using
\begin{eqnarray}
\bs{n}(\theta, \phi) = \frac{\bs{\sigma}_{\theta}(\theta, \phi) \times \bs{\sigma}_{\phi}(\theta, \phi)}{\| \bs{\sigma}_{\theta}(\theta, \phi) \times \bs{\sigma}_{\phi}(\theta, \phi) \|}. \label{equ:normal_vec_def}
\end{eqnarray}
The terms $\bs{\sigma}_{\theta\theta}$, $\bs{\sigma}_{\theta\phi}$, and $\bs{\sigma}_{\phi,\phi}$ are obtained by further differentiation from equation~\ref{eqn:sigmaTheta} and equation~\ref{eqn:sigmaPhi}.
We use the notation for the metric tensor $\mb{g} = \mathbf{I}$  interchangeably.  In practical calculations whenever we need to compute the action of the inverse metric tensor we do so through numerical linear algebra (Gaussian elimination with pivoting)~\cite{Trefethen1997,Strang1980}.  For notational convenience, we use the tensor notation for the metric tensor $g_{ij}$ and its inverse $g^{ij}$ which has the formal correspondence
\begin{eqnarray}
g_{ij} = \left[\mathbf{I}\right]_{i,j}, \hspace{0.3cm} g^{ij} = \left[ \mathbf{I}^{-1}\right]_{i,j}.
\end{eqnarray}
For the metric factor we also have that
\begin{eqnarray}
\sqrt{|g|} = \sqrt{\det(\mathbf{I})} = r^2\sqrt{r^{-2}r_{\theta}^2+(r^{-2}r_{\phi}^2+1)\sin^2(\phi)} = \| \vec{\sigma}_{\theta}(\theta, \phi) \times \vec{\sigma}_{\phi}(\theta, \phi) \|.
\label{met_fac_def}
\end{eqnarray}
To ensure accurate numerical calculations in each of the above expressions the appropriate coordinates either Chart A or Chart B are used to ensure sufficient distance from coordinate singularities at the poles.  To compute quantities associated with curvature of the manifold we use the Weingarten map~\cite{Pressley2001} which can be expressed as
\begin{eqnarray}
\mb{W} = -\mb{I}^{-1} \mb{II}.
\end{eqnarray}
To compute the Gaussian curvature $K$, we use
\begin{eqnarray}
K(\theta,\phi) = \det\left(\mb{W}(\theta,\phi)\right).
\end{eqnarray}
For further discussion of the differential geometry of manifolds see~\cite{Pressley2001,Abraham1988,SpivakDiffGeo1999}.

\section{Operators on Radial Manifolds}
\label{appendix:diff_op_r_manifold}
We compute operators on the surface by representing the geometry of the radial manifold as a finite spherical harmonics expansion $r(\theta,\phi) = \sum_{|k| \leq L} \hat{r}_k Y_k(\theta,\phi)$ where $r(\theta,\phi)$ represents the radial component of the radial manifold as in equation~\ref{equ:manifoldParam}.  In this manner we have an analytic representation of the geometry allowing for fundamental operators to be computed.  This involves a few common operators for which we give explicit expressions in coordinates that are used for this purpose.  Since there is no global non-singular coordinate system on the manifold surface, we ensure numerical accuracy by switching between two coordinate charts.  In chart $A$ we have coordinates $(\hat{\theta},\hat{\phi})$ with singularities at the north and south poles.  In chart $B$ we have coordinates $(\tilde{\theta},\tilde{\phi})$ having singularities at the east and west poles.  To avoid issues with singularities when seeking a value at a point $\mb{x}$, we evaluate expressions within each chart in the regions with $\pi/4 \leq \phi \leq 3\pi/4$ and $\pi/4 \leq \tilde{\phi} \leq 3\pi/4$.  We give all expressions with generic polar coordinates $(\theta,\phi)$ which we subsequently use in practice in our numerical calculations by choosing the appropriate chart $A$ or chart $B$.  

The scalar Laplace-Beltrami operator $\Delta_{LB} = -\bs{\delta} \mb{d}$ that acts on $0$-forms can be expressed in coordinates as
\begin{equation}
\Delta_{LB} = \frac{1}{\sqrt{|g|}}\partial_i \left(g^{ij}\sqrt{|g|} \partial_j \right).
\end{equation}
The $g_{ij}$ denotes the metric tensor, $g^{ij}$ the inverse metric tensor, and $|g|$ the determinant of the metric tensor as in Appendix~\ref{appendix:radialManifold}.  For the radial manifold when using the coordinates $(\theta,\phi)$ we find it useful to consider
\begin{eqnarray}
h_{ij} = \left(\sqrt{|g|}g^{ij}\right)\partial_{ij} + \left(\partial_{i}\sqrt{|g|}g^{ij}\right)\partial_j.
\end{eqnarray}
Here $\partial_1 = \partial_\theta$ and $\partial_2 = \partial_\phi$.  We obtain $\Delta_{LB} = (1/\sqrt{|g|}) \sum_{ij} h_{ij}$.  
The decomposition into $h_{ij}$ is particularly helpful since this makes more transparent the polar derivatives in $\phi$ which pose in practice the most challenges in numerical calculations.
This allows us to compute analytically many of the terms that arise involving the metric and exterior operators.  Using that the manifolds are two dimensional with coordinates $(\theta,\phi)$ we have from the formula for a two-by-two inverse matrix that
\begin{equation}
\label{equ_g_gij}
    \sqrt{|g|}g^{ij} = 
    \begin{cases}
       {g_{\phi \phi}}/{\sqrt{|g|}} & \text{if: $i = j = \theta$} \\
       {g_{\theta \theta}}/{\sqrt{|g|}} & \text{if: $i = j = \phi$} \\
       {-g_{\theta \phi}}/{\sqrt{|g|}} = {-g_{\phi \theta}}/{\sqrt{|g|}} & \text{if: $i \ne j$}.
    \end{cases}
\end{equation}
In the radial manifold case, we can compute each of these terms in the expression in equation~\ref{equ_g_gij} using the results from Appendix~\ref{appendix:radialManifold}.

The exterior derivatives can be expressed for a $0$-form $f$ and $1$-form $\bs{\alpha}$ as
\begin{eqnarray}
\mb{d}f &=& (\partial_{\theta}f) \mb{d}\theta + (\partial_{\phi} f) \mb{d}\phi = f_{\theta}\mb{d}\theta + f_{\phi}\mb{d}\phi \\
\mb{d}\alpha &=& (\partial_{\theta} \alpha_{\phi} - \partial_{\phi} \alpha_{\theta}) \mb{d}\theta \wedge \mb{d}\phi.
\end{eqnarray}
The generalized curl on the radial manifold of a $0$-form and $1$-form can be expressed as
\begin{eqnarray}
-\star \mb{d}f & = & \mbox{curl}_\mathcal{M}(f) = 
\sqrt{|g|}(f_{\theta}g^{\theta \phi} + f_{ \phi} g^{\phi \phi})\mb{d}\theta - \sqrt{|g|}(f_{\theta}g^{\theta \theta} + f_{\phi}g^{\phi \theta})\mb{d}\phi \\
-\star \mb{d} \alpha & = & \mbox{curl}_\mathcal{M}(\bs{\alpha}) = \frac{\partial_{\phi} \alpha_{\theta} - \partial_{\theta} \alpha_{\phi} }{\sqrt{|g|}}.
\end{eqnarray}
In this notation we have taken the conventions that
$f_j = \partial_{x^j} f$ and $\alpha_j$ such that $\bs{\alpha} = \alpha_j \mb{d}{x}^j$ where $j \in \{\theta,\phi\}$. 
The isomorphisms $\sharp$ and $\flat$ between vectors and co-vectors can be expressed explicitly as
\begin{eqnarray}
\mb{v}^{\flat} &=& (v^{\theta}\bs{\sigma}_{\theta} + v^{\phi}\bs{\sigma}_{\phi})^{\flat}\\
\nonumber
 &=& v^{\theta}g_{\theta \theta}\mb{d}\theta + v^{\theta}g_{\theta \phi}\mb{d}\phi + v^{\phi}g_{\phi \theta}\mb{d}\theta + v^{\phi}g_{\phi \phi}\mb{d}\phi \\
 \nonumber
  &=& (v^{\theta}g_{\theta \theta} + v^{\phi}g_{\phi \theta})\mb{d}\theta + (v^{\theta}g_{\theta \phi} + v^{\phi}g_{\phi \phi})\mb{d}\phi \\ 
\nonumber
\\
(\bs{\alpha})^{\sharp} &=& (\alpha_{\theta}\mb{d}\theta + \alpha_{\phi}\mb{d}\phi)^{\sharp} \\
\nonumber
 &=& \alpha_{\theta}g^{\theta \theta}\bs{\sigma}_{\theta} + \alpha_{\theta}g^{\theta \phi}\bs{\sigma}_{\phi} + \alpha_{\phi}g^{\phi \theta}\bs{\sigma}_{\theta} + \alpha_{\phi}g^{\phi \phi}\bs{\sigma}_{\phi} \\
 \nonumber
 &=& (\alpha_{\theta}g^{\theta \theta} + \alpha_{\phi}g^{\phi \theta})\bs{\sigma}_{\theta} + (\alpha_{\theta}g^{\theta \phi} + \alpha_{\phi}g^{\phi \phi})\bs{\sigma}_{\phi}
\end{eqnarray}
We use the notational conventions here that for the embedding map $\bs{\sigma}$ we have $\bs{\sigma}_\theta = \partial_{\theta}$ and $\bs{\sigma}_\phi = \partial_{\phi}$ as in Appendix~\ref{appendix:radialManifold}.  Combining the above equations we can express the generalized curl as
\begin{eqnarray}
\label{equ_gen_curl_0form}
(-\star \mb{d}f)^{\sharp} &=& ([\sqrt{|g|}(f_{\theta}g^{\theta \phi} + f_{ \phi} g^{\phi \phi})]g^{\theta \theta} + [- \sqrt{|g|}(f_{\theta}g^{\theta \theta} + f_{\phi}g^{\phi \theta})]g^{\phi \theta})\bs{\sigma}_{\theta}\\ 
\nonumber
&+& ([\sqrt{|g|}(f_{\theta}g^{\theta \phi} + f_{ \phi} g^{\phi \phi})]g^{\theta \phi} + [- \sqrt{|g|}(f_{\theta}g^{\theta \theta} + f_{\phi}g^{\phi \theta})]g^{\phi \phi})\bs{\sigma}_{\phi} \\
\nonumber
&=& \frac{f_{\phi}}{\sqrt{|g|}} \bs{\sigma}_{\theta} - \frac{f_{\theta}}{\sqrt{|g|}} \bs{\sigma}_{\phi} \\ 
\nonumber
\\
\label{equ_gen_curl_1form}
-\star \mb{d}\mb{v}^{\flat} &=& -\frac{\partial_{\phi}(v^{\theta}g_{\theta \theta} + v^{\phi}g_{\phi \theta}) - \partial_{\theta} (v^{\theta}g_{\theta \phi} + v^{\phi}g_{\phi \phi}) }{\sqrt{|g|}}.
\end{eqnarray}

We also mention that the velocity field of the hydrodynamic flows $\mb{v}$ is recovered from the vector potential $\Phi$ as $\mb{v}^{\flat} = -\star \mb{d} \Phi$.  We obtain the velocity field $\mb{v} = \mb{v}^{\sharp} = \left(-\star \mb{d} \Phi\right)^{\sharp}$ using equation~\ref{equ_gen_curl_0form}.  Similarly from the force density $\mb{b}$ acting on the fluid, we obtain the data $-\star\mb{d}\mb{b}^{\flat}$ for the vector potential formulation of the hydrodynamics in equation~\ref{equ_Phi_fluid} using equation~\ref{equ_gen_curl_1form}.  Additional details and discussions of these operators also can be found in our related papers~\cite{AtzbergerSoftMatter2016,AtzbergerGross2017} and in~\cite{Pressley2001,Abraham1988, SpivakDiffGeo1999}.

\section{Spherical Harmonics}
\label{appendix:sphericalHarmonics}
We represent fields on the radial manifolds using spherical harmonics expansions.  We give a brief overview of spherical harmonics and how they are used in our numerical methods.  A more detailed discussion of spherical harmonics can be found in~\cite{HanBookSphericalHarmonics2010}. More details on how we use spherical harmonics in our numerical methods to compute exterior calculus operators can be found in our papers~\cite{AtzbergerGross2017,AtzbergerSoftMatter2016} and in Appendix~\ref{appendix:radialManifold}.

The spherical harmonics are the eigenfunctions of the Laplace-Beltrami operator $\Delta_{LB} = -\bs{\delta}\mb{d}$ on the sphere.  In spherical coordinates they can be expressed as
\begin{equation}
\label{equ:sphericalHarmonics}
Y^m_n(\theta, \phi) = \sqrt{\frac{(2n+1)(n-m)!}{4\pi(n+m)!}}P^m_n\left(\cos(\phi)\right) \exp\left({im\theta}\right).
\end{equation}
The $m$ denotes the order and $n$ the degree for $n \ge 0$ and $m \in \{-n, \dots, n\}$.  The $P^m_n$ denote the \textit{Associated Legendre Polynomials}.  We denote by $\theta$ the azimuthal angle and by $\phi$ the polar angle of the spherical coordinates~\cite{HanBookSphericalHarmonics2010}.

We work with real-valued functions and use that modes are self-conjugate in the sense $Y^m_n = \overline{Y^{-m}_n}$.  We can express spherical harmonic modes as
\begin{equation}
Y^m_n(\theta, \phi) = X^m_n(\theta, \phi)+ iZ^m_n(\theta, \phi).
\end{equation}
The $X_n^m$ and $Z_n^m$ denote the real and imaginary parts.  We use this splitting in our numerical methods to construct a purely real set of basis functions on the unit sphere with maximum degree $N$ which consists of $(N+1)^2$ basis elements. For the case $N=2$ we have the basis elements  
\begin{eqnarray}
\tilde{Y}_1 = Y^0_0,\hspace{0.2cm} 
\tilde{Y}_2 = Z^1_1,\hspace{0.2cm} 
\tilde{Y}_3 = Y^0_1,\hspace{0.2cm} 
\tilde{Y}_4 = X^1_1,\hspace{0.2cm} 
\tilde{Y}_5 = Z^2_2,\hspace{0.2cm} 
\tilde{Y}_6 = Z^1_2,\hspace{0.2cm} 
\tilde{Y}_7 = Y^0_2,\hspace{0.2cm} 
\tilde{Y}_8 = X^1_2,\hspace{0.2cm} 
\tilde{Y}_9 = X^2_2.\hspace{0.2cm} 
\end{eqnarray}
Similar conventions are used for the basis for the other values of $N$.  We take final basis elements $Y_i$ that are normalized as $Y_i = \tilde{Y}_i/\sqrt{\langle \tilde{Y}_i, \tilde{Y}_i \rangle}$.

Derivatives are used within our finite expansions by evaluating analytic formulas whenever possible for the spherical harmonics in order to try to minimize approximation error~\cite{HanBookSphericalHarmonics2010}.  Approximation errors are incurred when sampling the values of expressions involving these derivatives at the Lebedev nodes and when performing quadratures.  The derivative of the spherical harmonics in the azimuthal coordinate $\theta$ is given by
\begin{equation*}
\partial_\theta Y^m_n(\theta, \phi) 
= \partial_\theta \sqrt{\frac{(2n+1)(n-m)!}{4\pi(n+m)!}}P^m_n(\cos(\phi)) \exp\left({im\theta}\right)
= imY^m_n\left(\theta, \phi\right).
\end{equation*}
We see this has the useful feature that the derivative in $\theta$ of a spherical harmonic of degree $n$ is again a spherical harmonic of degree $n$.  As a consequence, we have in our numerics that this derivative can be represented in our finite basis.  This allows us to avoid additional $L^2$  projections allowing for computation of the derivative in $\theta$ without incurring an approximation error.  The derivative of the spherical harmonics in the polar angle $\phi$ is given by
\begin{equation}
\label{equ:derivPhi}
\partial_\phi Y^m_n(\theta, \phi) 
= m \cot(\phi)Y^m_n(\theta, \phi)+\sqrt{(n-m)(n+m+1)}\exp\left({-i\theta}\right) Y^{m+1}_n(\theta, \phi).
\end{equation}
We see that unlike derivatives in $\theta$ the derivative in $\phi$ can not be represented in general in terms of a finite expansion of spherical harmonics.  In our numerics, we use the expression in equation~\ref{equ:derivPhi} for $\partial_\phi Y^m_n(\theta, \phi)$ when we need to compute values at the Lebedev quadrature nodes.  These analytic results provide a convenient way to compute derivatives of differential forms following the approach discussed in our prior paper~\cite{AtzbergerGross2017}.  By using these analytic expressions, we have that the subsequent hyperinterpolation of the resulting expressions are where the approximation errors are primarily incurred.  Throughout our discussions to simplify the notation we use the convention that $Y^{m}_n = 0$ when $m \geq n+1$.  Further discussion of spherical harmonics can be found~\cite{HanBookSphericalHarmonics2010}.  Further discussions about how we use the spherical harmonics in our numerical calculations of exterior calculus operators also can be found in our paper~\cite{AtzbergerGross2017}.

\end{document}